\documentclass[pdflatex,sn-mathphys-num]{sn-jnl}

\usepackage{graphicx}%
\usepackage{multirow}%
\usepackage{amsmath,amssymb,amsfonts}%
\usepackage{amsthm}%
\usepackage{mathrsfs}%
\usepackage[title]{appendix}%
\usepackage{xcolor}%
\usepackage{textcomp}%
\usepackage{manyfoot}%
\usepackage{booktabs}%
\usepackage{algorithm}%
\usepackage{algorithmicx}%
\usepackage{algpseudocode}%
\usepackage{listings}%
\usepackage{setspace}
\usepackage{braket}
\usepackage{xcolor}
\definecolor{myorange}{RGB}{255,149,0}
\definecolor{mygreen}{RGB}{0,185,69}
\definecolor{myblue}{RGB}{12,93,165}
\usepackage{tabularx}

\theoremstyle{thmstyleone}%
\theoremstyle{thmstyletwo}%

\theoremstyle{thmstylethree}%

\begin{document}



\title[Article Title]{Four- and six-photon stimulated Raman transitions for coherent qubit and qu\textit{d}it operations}

\author*[1]{\fnm{Gabriel} \sur{Gregory}}\email{ggregory@uoregon.edu}
\author[1]{\fnm{Evan} \sur{Ritchie}}
\author[1]{\fnm{Alex} \sur{Quinn}}
\author[1]{\fnm{Sean} \sur{Brudney}}
\author[1]{\fnm{David} \sur{Allcock}}
\author[1]{\fnm{David} \sur{Wineland}}
\author*[1]{\fnm{Jameson} \sur{O'Reilly}}\email{joreilly@uoregon.edu}

\affil[1]{\orgdiv{Department of Physics}, \orgname{University of Oregon}, \orgaddress{\street{1371 E 13th Ave}, \city{Eugene}, \postcode{97403}, \state{OR}, \country{USA}}}


\abstract{Quantum computers are typically composed of an array of two-level systems, or qubits, encoded in some information carrier, such as an electron, photon, or quantized circuit. The size of this array is restricted by finite access to resources like laser power, cooling capacity, and control lines for trapping and manipulation. Under these constraints, the system’s processing power can be increased by using more energy levels per information carrier, but common techniques for qubit control provide only limited connectivity between these additional states. We experimentally demonstrate transitions between electronic angular momentum states with a difference in magnetic quantum numbers $\Delta \mathrm{m_J} = $ 3, 4, and 5 via resonant four- and six-photon stimulated Raman transitions in a single trapped atom. Derivation of analytic formulas for the corresponding Rabi frequencies, which are verified experimentally, follows the standard treatment of two-photon transitions including the adiabatic elimination of intermediate states. Finally, we discuss pathways to increase the observed multi-photon transition fidelities to $>99.99\%$, providing a tool for efficient, high-fidelity control of high-dimensional qu\textit{d}its and single-atom logical qubits.}


\maketitle

\section*{Introduction}\label{sec1}

Two-photon stimulated Raman transitions are a well-developed tool for unitary control over qubits encoded in trapped ions~\cite{daveRaman}, neutral atoms~\cite{Yavuz2006,Jones2007}, solid state systems~\cite{Goldman2020,Yale2013}, and molecules~\cite{molecularRaman}. They can be deployed in large-scale systems with absolute phase stability in space and time~\cite{Inlek2014} and allow for individual addressing of qubits~\cite{Hffner2005}. Their implementation has enabled sub-\textmu s quantum logic gate operations~\cite{schafer_fast_2018} and single and two-qubit gate fidelities in trapped ions exceeding $99.99\%$ and $99.9\%$, respectively~\cite{Gaebler2016,ballance2016}. In addition to computing, two-photon stimulated Raman transitions have also found uses in matter-wave interferometry~\cite{Kasevich1991,Muller2008} and quantum simulation~\cite{Monroe2021}. These transitions typically consist of simultaneous absorption and emission of single photons via electric dipole transitions and thus are limited to coupling quantum states with a difference of at most two units of angular momentum. 

There has been growing interest in encoding quantum information within the higher-dimensional Hilbert spaces native to atomic systems~\cite{hrmo2023native,jia2024architecture,low2025control,Kiktenko2025}. So-called qu$d$it encodings (with dimension $d>2$) have been shown to simplify circuits~\cite{Lanyon2009} and tomography~\cite{kueng2022} and enable efficient quantum simulation of higher-dimensional physical systems~\cite{Zoller2022,senkosimulation}. Numerical simulations have also shown more efficient error correction using qudits~\cite{Watson2015}, with experimental implementations of codes embedded in single ion qu$d$its recently demonstrating better-than-physical performance~\cite{debry2026error,Li2025}. While some implementations have maximized dimension by encoding computational states in both metastable and ground manifolds ~\cite{jia2024architecture, hrmo2023native,low2025control}, metastable-only encodings enable the conversion of almost all leakages out of the qu$d$it subspace into erasure errors~\cite{alexMgate,shi2025long} while still providing a large Hilbert space~\cite{grover_qudit}. This feature has been shown to reduce the overhead of fault tolerant quantum error correction in numerical simulations~\cite{neutralErasure,erasureConversion}. In addition, metastable encodings are compatible with same-species sympathetic cooling and mid-circuit readout~\cite{Allcock2021,yang2022realizing}, thus reducing experimental complexity. 

Large metastable qu$d$it encodings inevitably contain state pairs with $|\Delta \mathrm{m}| > 2$ or $|\Delta\mathrm{F}|>2$ that consequently cannot be directly coupled with dipole-based two-photon processes~\cite{waterlooqudits}, so realizing some unitaries will require longer sequences of pulses coupling to intermediate qu\textit{d}it states. Specific examples include logical operations in \AE~\cite{AEcode} and other single-atom quantum error correction codes~\cite{spincats}. This can be mitigated to some extent by driving sequences of single-photon electric quadrupole transitions, but these are temperature sensitive, reliant on narrow-linewidth lasers, and not straightforwardly compatible with same-species sympathetic cooling. Multi-photon quadrupole transitions could in principle bypass some of these limitations but may be slow and to the best of our knowledge have not been demonstrated.

To bridge this gap, we propose to use $2n$-photon stimulated Raman transitions to directly couple states separated by up to $2n$ units of angular momentum. Four-photon Raman processes have previously been observed in ensembles of neutral atoms~\cite{4photonBEC,chang4photon} and carbon nanotubes~\cite{bunkinnanotubes}, but not in individual atomic systems with high-fidelity state transfer and readout. In one study, fifty-photon Raman transitions were demonstrated with high efficiency for a $d=2$ system in a gas of Rb atoms~\cite{Hansch2001}. More common examples of multi-photon processes include four-wave mixing in cold atom ensembles~\cite{walker2008}, which generally involves four separate manifolds, and Bragg diffraction for matter wave interferometers~\cite{Muller2008}, which does not change the internal state. Neither technique, to the best of our knowledge, has been directly incorporated into quantum information processing. Four-photon ac Stark shifts have been measured in trapped ions~\cite{monroe4starkshift,jung2025ion}, but a detailed study of higher order processes driving coherent population transfer within an isolated manifold has not yet been put forth.

In this Article, we present a theoretical treatment and experimental verification of stimulated Raman transitions driven by four- and six-photon processes capable of directly coupling state pairs with $|\Delta \mathrm{m_J}|>2$. We derive analytic formulas for the corresponding Rabi frequencies and verify them for specific transitions with $|\Delta \mathrm{m_J}|=3$ and $|\Delta \mathrm{m_J}|=4$ in the $\mathrm{D_{5/2}}$ metastable manifold of a single trapped $^{40}$Ca$^+$ ion. We achieve four- and six-photon $\pi$-pulse transfer fidelities of $96(1)$\% and $78(4)$\% and discuss the path towards decreasing the infidelities below $10^{-4}$. Finally, we drive select $\Delta \mathrm{m_J}>2$ transitions that enable full state-to-state connectivity of a metastable qu$d$it in the $\mathrm{D_{5/2}}$ manifold, providing a potential pathway towards more efficient Raman-based qu\textit{d}it control.

\section*{Analytic Rabi frequency predictions}\label{sec:analytics}
\begin{figure}[h]
    \centering
    \hspace*{-0.0in}
    \includegraphics[width=0.7\columnwidth]{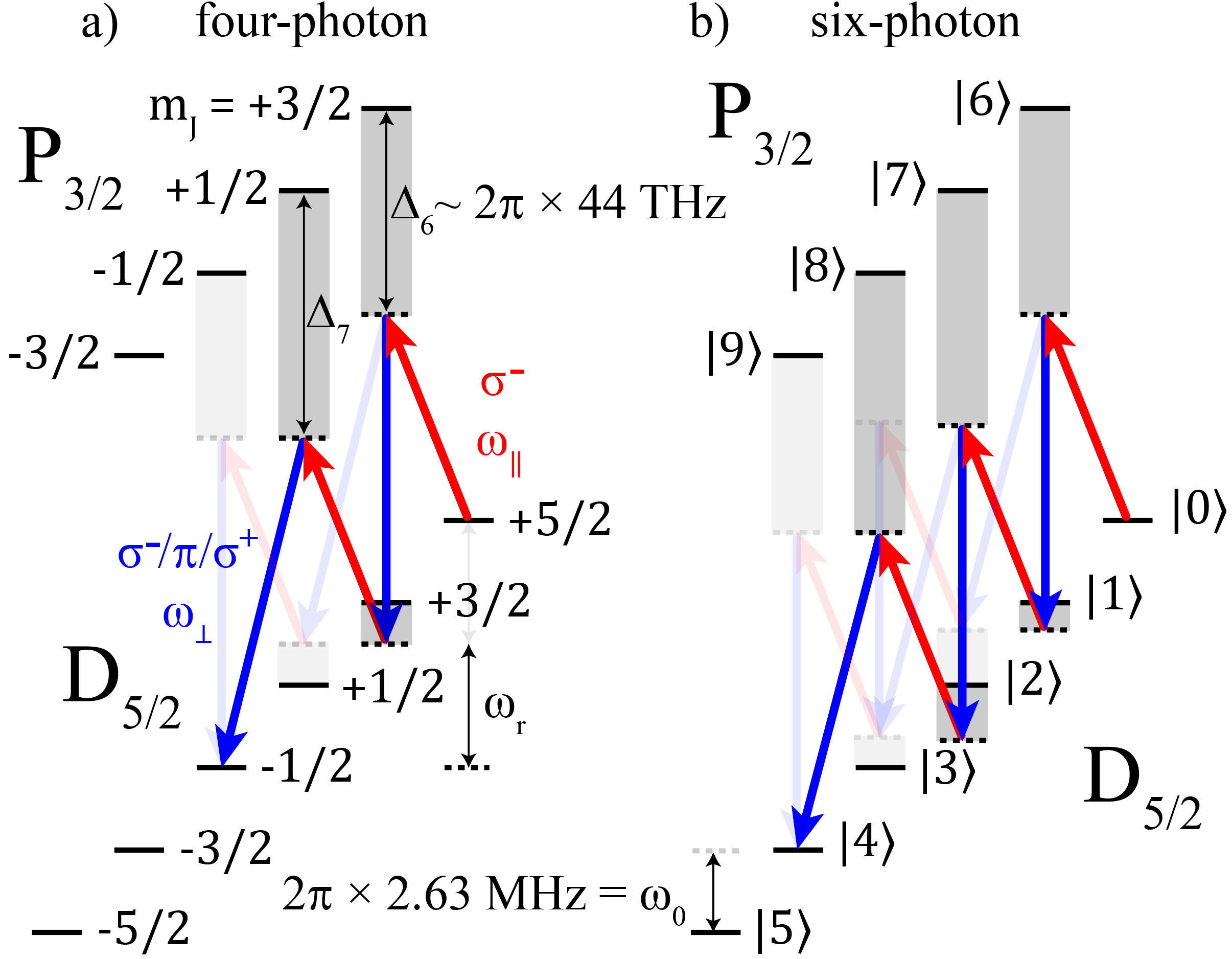}
    \caption{\textbf{\textbar\ Four- and six-photon pathways.} \textbf{a,} Illustration of four-photon transition pathways between $\ket{0}\equiv\ket{\mathrm{m_J} = +5/2}$ and $\ket{3}\equiv\ket{\mathrm{m_J} = -1/2}$ in the $\mathrm{D_{5/2}}$ manifold of $^{40}$Ca$^+$. They are driven by two 976\,nm beams with $\sigma^-$ (red) and equal components of $\sigma^+$, $\pi$ and $\sigma^-$ (blue) polarizations and a relative detuning $\omega_r$. \textbf{b,} Six-photon transition between $\ket{0}$ and $\ket{4}$. Solid and transparent lines represent (three of the five) interfering pathways that drive the multi-photon transitions, with solid lines highlighted as an example pathway. Energies are not to scale.}
    \label{fig:paths_flopping}
\end{figure}

For concreteness, we present derivations of four- and six-photon processes within the $\mathrm{^3D_{5/2}}$ manifold of $^{40}$Ca$^+$, but the underlying physics can be straight-forwardly extended to other systems. The Zeeman structure of the metastable $\mathrm{D_{5/2}}$ and short-lived $\mathrm{P_{3/2}}$ manifolds (lifetimes of 1.168(7)\,s~\cite{Barton2025} and  6.64(4)\,ns~\cite{Meir2020}, respectively) are shown in Fig.~\ref{fig:paths_flopping}, with metastable levels split by $\omega_0=2\pi\times2.63$\,MHz due to a quantization field of $\sim1.56$\,G. 

We consider dynamics driven by two beams far-detuned by $\Delta$ from the 854\,nm $\mathrm{D_{5/2}}\leftrightarrow \mathrm{P_{3/2}}$ resonance. One beam ($R_{\parallel}$) we assume to be aligned with the magnetic field with nominally pure $\sigma^-$ polarization, frequency $\omega_{\parallel}$, and power $P_\parallel$ (red in Fig.~\ref{fig:paths_flopping}). The other beam ($R_{\perp}$) propagates orthogonally to the magnetic field with frequency $\omega_{\perp}$ and power $P_\perp$ and contains equal components of $\sigma^+$, $\pi$ and $\sigma^-$ polarized light (blue in Fig.~\ref{fig:paths_flopping}). For an \textit{n}-photon transition between states $\ket{i}$ and $\ket{j}$, the relative beam detuning $\omega_r \equiv\omega_\perp-\omega_\parallel$ is set to satisfy the resonance condition
\begin{align}
   \omega_r = \frac{2\left|E_i - E_j\right|}{n\hbar}.
    \label{resonance}
\end{align}

The interaction of a laser field with states $\ket{i}$ and $\ket{j}$ is expressed by the $ij$'th component of the interaction Hamiltonian
\begin{align}
    H_{ij}=-\frac{\hbar}{2}\Omega_{ij}e^{-i\omega t}
\end{align}
where $\Omega_{ij}$ is the complex single-photon Rabi frequency and $\omega$ is the laser frequency. Our approach (Methods~\ref{sec:methods_derivation} for an overview and Appendix~\ref{appendix:fourphotonderivation} for details) mirrors the standard treatment of the two-photon Raman Rabi frequency~\cite{daveRaman}, wherein state amplitude equations of motion are generated via the Schr\"odinger equation. The single-beam detuning $\Delta$ is taken to be much larger than the time derivative of the excited state's corresponding complex amplitude. Under that assumption, the state may be adiabatically eliminated, uncoupling the equations of motion and reducing the problem to an effective two-state system. The key new approximation that we make in this treatment is that the MHz-scale detuning from intermediate states within the $\mathrm{D_{5/2}}$ is enough to adiabatically eliminate those states as well (Appendix~\ref{appendix:numericalsims}).


Using the state labels defined in Fig.~\ref{fig:paths_flopping}, we thus find the four- and six-photon Rabi frequencies

\begin{equation}
\label{4photon_rabi}
\mathbf{\Omega_{03}^{(4)}}=\frac{\Omega^*_{06}}{8\Delta^2}\left(\frac{\Omega_{16}\Omega^*_{17}\Omega_{37}}{\omega_r-\omega_{01}}-\frac{\Omega_{26}\Omega^*_{28}\Omega_{38}}{\omega_r-\omega_{23}}\right)
\end{equation}
\begin{equation}
\label{6photon_rabi}
\begin{split}
\mathbf{\Omega_{04}^{(6)}}=\frac{\Omega^*_{06}}{32\Delta^3}&\left(\frac{\Omega_{16}\Omega^*_{17}\Omega_{27}\Omega^*_{28}\Omega_{48}}{(\omega_r-\omega_{01})(2\omega_{r}-\omega_{02})}\right.\\ &- \frac{\Omega_{16}\Omega^*_{17}\Omega_{37}\Omega^*_{39}\Omega_{49}}{(\omega_{r}-\omega_{01})(\omega_{03}-2\omega_{r})}\\ &+\frac{\Omega_{26}\Omega^*_{28}\Omega_{38}\Omega^*_{39}\Omega_{49}}{(\omega_{02}-\omega_{r})(\omega_{03}-2\omega_{r})}\\ &- \frac{\Omega_{26}\Omega^*_{28}\Omega_{28}\Omega^*_{28}\Omega_{48}}{(\omega_{02}-\omega_r)(2\omega_{r} - \omega_{02})}\\ &+ \left. \frac{\Omega_{06}\Omega^*_{06}\Omega_{26}\Omega^*_{28}\Omega_{48}}{\omega_{r}(2\omega_{r} - \omega_{02})}\right)
\end{split}
\end{equation}

\noindent where $\omega_{ij}$ is the angular frequency splitting between states $\ket{i}$ and $\ket{j}$ including corrections due to ac Stark shifts induced by the $R_\parallel$ and $R_\perp$ beams. Most of the terms in Equations~\eqref{4photon_rabi} and~\eqref{6photon_rabi} are represented as pathways in Fig.~\ref{fig:paths_flopping}. We find convergence between these equations and numerical simulations of the full $\mathrm{D_{5/2}}$ interaction Hamiltonian detailed in Appendix~\ref{appendix:numericalsims}. In general, ignoring interference between different pathways, the two-, four-, and six-photon transition Rabi frequencies scale as
\begin{align}
    \label{2photon}
    \mathbf{\Omega}^{(2-photon)} &\propto \frac{\Omega^2}{\Delta}\\
    \label{4photon}
    \mathbf{\Omega}^{(4-photon)} &\propto \frac{\Omega^4}{\Delta^2\omega_0}\\
    \label{6photon}
    \mathbf{\Omega}^{(6-photon)} &\propto \frac{\Omega^6}{\Delta^3\omega^2_0}
\end{align}
which corresponds to $(P_\perp P_\parallel)^{n/4}$ scaling with the beam powers for $n$-photon transitions.

\section*{Spectroscopy and power scaling}
\label{sec:measurements}

\begin{figure}[h]
\centering
    \hspace*{-0.0in}
    \includegraphics[width = 0.7\columnwidth]{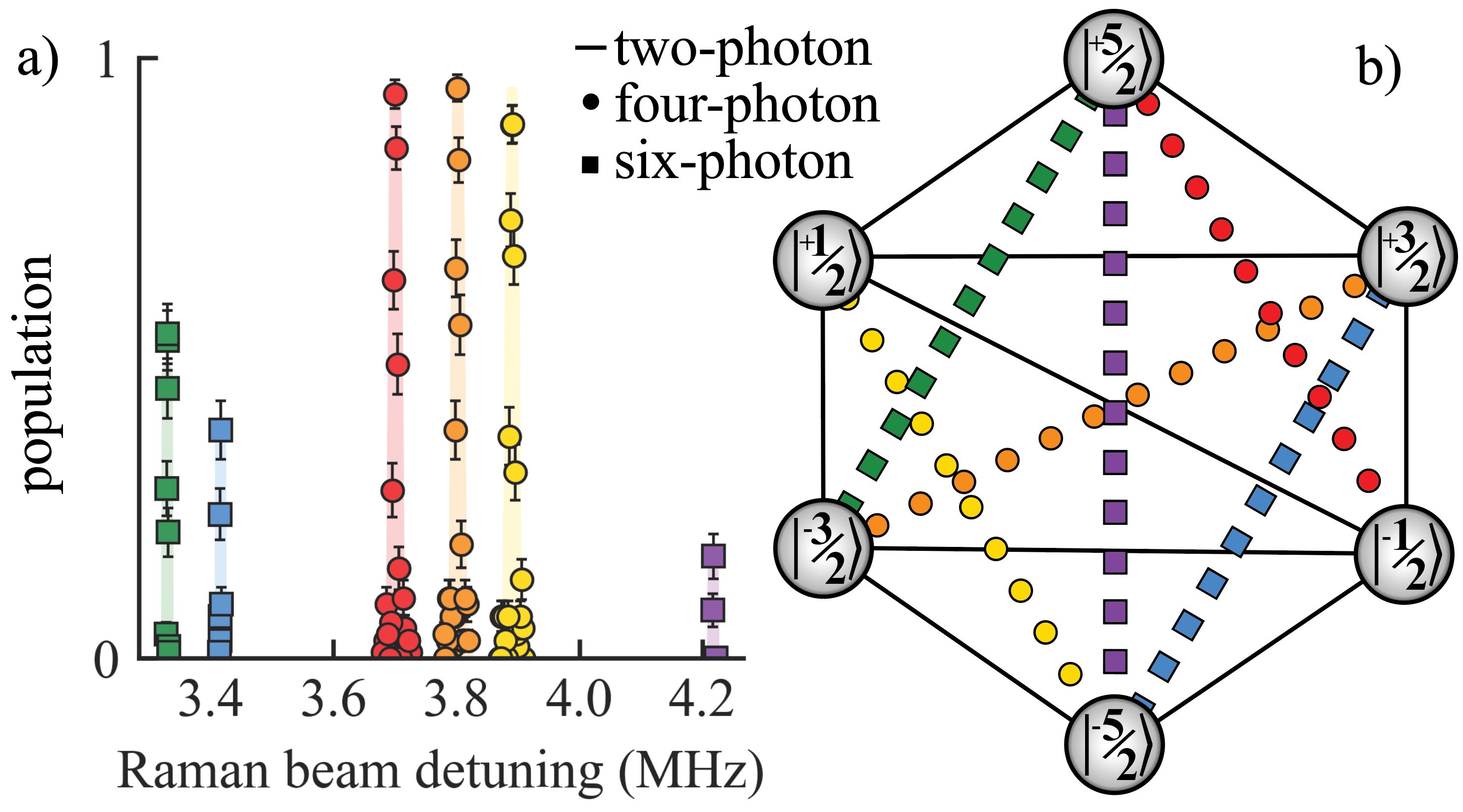}
    \caption{\textbf{\textbar\ All-to-all qu$d$it connectivity. }\textbf{a,} Rabi spectroscopy of transitions within $\mathrm{D_{5/2}}$ with $\Delta \mathrm{m_J} \geq 3$ as driven by four (circles) and six (squares) photon processes. \textbf{b,} Diagram illustrating full direct unitary connectivity in the $\mathrm{D_{5/2}}$ manifold enabled by four- and six-photon transitions.}
\label{fig:spectroscopy}
\end{figure}

To validate these predictions, we performed experiments with a single trapped and laser-cooled $^{40}$Ca$^+$ ion driven by 976\,nm stimulated Raman transition laser beams with $\Delta=-2\pi\times44$\,THz. Our trap and Raman laser system are described in Ref.~\cite{alexMgate}. Each 976\,nm beam is focused to a $\sim$30\,\textmu m waist at the ion and they have maximum powers of $P_\parallel=195$\,mW and $P_\perp=180$\,mW. Using state preparation and readout techniques described in Methods~\ref{sec:methods_spam}, we recover full population information from all six $\mathrm{D_{5/2}}$ sublevels after every experiment, enabling the direct validation of population driving between states with $\Delta\mathrm{m}>2$. As depicted in Fig.~\ref{fig:spectroscopy}a, we performed spectroscopy of four- and six-photon transitions between every pair of $\mathrm{D_{5/2}}$ sublevels that cannot be bridged by a two-photon transition, enabling full state-to-state connectivity within a $d=6$ qu\textit{d}it. Spectral isolation of each transition is provided by $R_\parallel$-induced ac Stark shifts. This connectivity is illustrated in Fig.~\ref{fig:spectroscopy}b, with transitions driven by two, four, and six photon processes marked with lines, circles, and squares, respectively.

\begin{figure*}[!t]
    \centering
    \includegraphics[width = \textwidth]{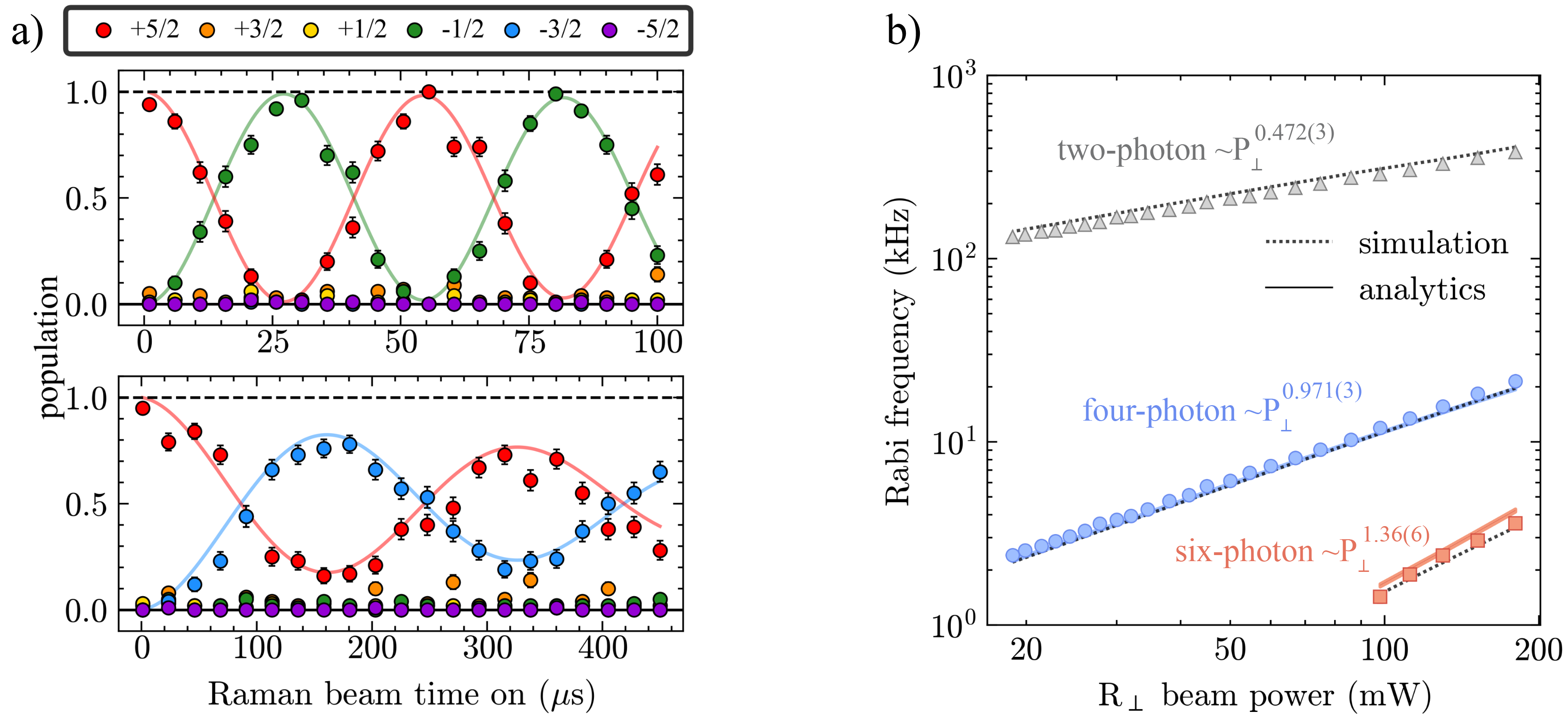}
    \caption{\textbf{\textbar\ Rabi oscillations and beam power scaling. }\textbf{a,} (Top) Rabi-flopping driven by the resonant four-photon process displayed in Fig.~\ref{fig:paths_flopping}a. (Bottom) Rabi flopping driven by the six-photon resonant transition shown in Fig.~\ref{fig:paths_flopping}b. \textbf{b,}  Two- (gray), four- (blue) and six-photon (red) transition Rabi frequencies as a function of $P_\perp$, with $P_\parallel$ held fixed at 195\,mW. Blue and red solid lines correspond to the predictions of Equations~\eqref{4photon_rabi} and~\eqref{6photon_rabi}, respectively, plus corrections for $\mathrm{F}$-state couplings, counter-rotating pathways, higher order corrections, and polarization impurities (see Appendices~\ref{appendix:calibrations} and~\ref{appendix:analytic_corrections}). Shaded regions represent 68\% confidence intervals from fits to beam intensities and polarizations. Dashed lines are predictions from numerical simulations.}
\label{fig:powerscaling}
\end{figure*}

Next, we studied the laser beam power dependence of Eqs.~\eqref{4photon_rabi} and~\eqref{6photon_rabi}, corresponding to the red and green transitions in Fig.~\ref{fig:spectroscopy}, respectively. To avoid excessive ac Stark shift variations, the power in the $R_\parallel$ beam was held constant at 195\,mW while the power in the $R_\perp$ beam was varied from 20 - 180\,mW to probe dynamics at different timescales. At each beam power setting, we performed Rabi spectroscopy of the two transitions and then measured resonant driving as shown in Fig.~\ref{fig:powerscaling}a with 195 and 152\,mW in the $R_\parallel$ and $R_\perp$ beams, respectively. To extract a Rabi frequency from the damped oscillations, we fit to a magnetic field noise model described in Appendix~\ref{appendix:decoherence}. 

We calibrate the intensity in each polarization component of the $R_{\parallel}$ and $R_{\perp}$ beams by performing auxiliary measurements including the two-photon $\mathrm{m_J}=+5/2\leftrightarrow+3/2$ transition Rabi frequency at different $R_\perp$ beam powers (see Appendix~\ref{appendix:calibrations}). We find good agreement between the measured four- and six-photon Rabi frequencies, the converging analytic predictions given by Equations~\eqref{4photon_rabi} and~\eqref{6photon_rabi}, and numerical simulations, all shown in Fig.~\ref{fig:powerscaling}, including small corrections due to polarization impurities, couplings to $\mathrm{F}$ manifold states, counter-rotating terms, and higher order corrections (see Appendix~\ref{appendix:analytic_corrections}). The remaining discrepancy between the measured and predicted four-photon Rabi frequencies is most likely caused by a miscalibration of the beam parameters, potentially due to an uncompensated ac Zeeman shift.

\section*{Roadmap to high-fidelity operations}
\label{sec:discussion}

\begin{figure}[h]
\centering
    \hspace*{-0.0in}
    \includegraphics[width = 0.7\columnwidth]{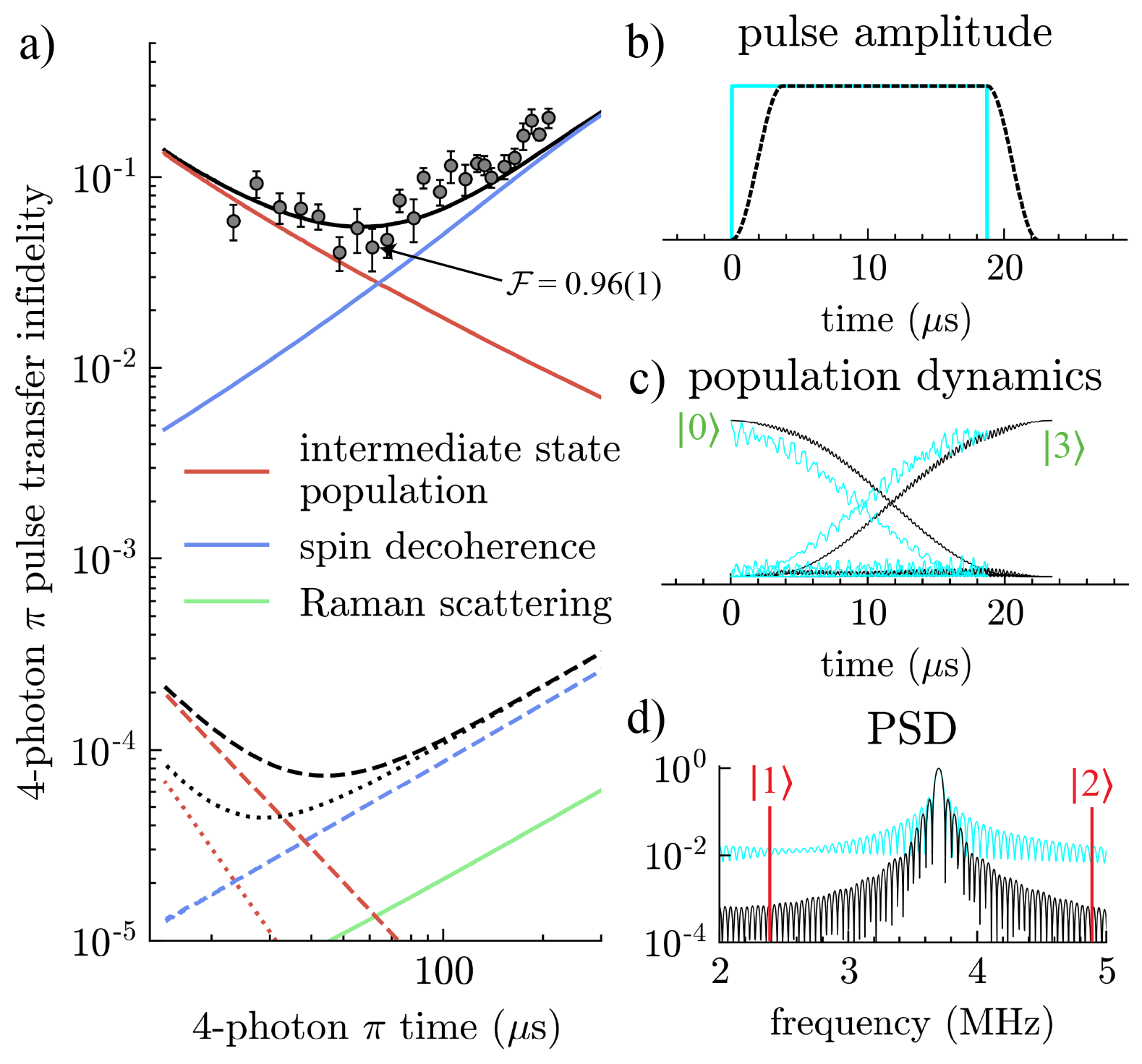}
    \caption{\textbf{\textbar\ Measured and achievable fidelity. }\textbf{a,} Four-photon $\pi$ pulse transfer infidelity at different four-photon $\pi$ times, corresponding to different $R_\perp$ beam powers, with expected contributions due to intermediate state population (solid red), spin decoherence (solid blue), and spontaneous Raman scatter error (solid green). Solid black line represents the sum of each expected contribution. Dashed blue line is projected spin dephasing infidelity with coherence demonstrated in~\cite{coherenceImprovement}. Dashed and dotted red lines are the numerically-simulated maximum leakage to intermediate states with the $R_\perp$ beam amplitude ramped with $\sin^2$ and $\sin^4$ envelopes respectively. \textbf{b,} Two-beam pulse envelopes with $R_\perp$ amplitude without pulse shaping (black) and shaped with a $\sin^2$ ramp (cyan). \textbf{c,} Numerically simulated resonant population dynamics driven by pulses shown in b). \textbf{d,} Power spectral density of pulses in b).} 
\label{fig:fidelity_and_int_states}
\end{figure}

The contrast of the four-photon Rabi flops, corresponding to a $\pi$ pulse transfer fidelity from $\ket{0}$ to $\ket{3}$, is limited by two dominant mechanisms presented in Fig.~\ref{fig:fidelity_and_int_states}. At low beam powers, ambient magnetic field fluctuations significantly damp the Rabi oscillations, while at high beam powers we observe measurable population of the intermediate states with fast, MHz-scale oscillations. We numerically simulate resonant dynamics of the six-level system and set the maximum total population in the intermediate states as an upper bound on the infidelity. Experimentally, we achieve a maximal four-photon $\pi$ pulse transfer fidelity of 96(1)\% at 48.6(3)\,\textmu s, with fidelity limitations in the fast and slow $\pi$ time regimes in line with predictions considering the two dominant error sources.

These results can be greatly improved, potentially to a level competitive with the best two-photon stimulated Raman gates~\cite{helios}, by implementing well-known techniques including magnetic field stabilization~\cite{coherenceImprovement} and pulse shaping (see Appendix~\ref{appendix:highfid}).
Square pulses, like we used in these experiments, have significant frequency components at the intermediate state two-photon resonances. If we instead ramp the beam intensities on and off, with sin$^2$ and sin$^4$ ramps given as an example in Fig.~\ref{fig:fidelity_and_int_states}, the power spectral density (PSD) at the intermediate state frequencies is reduced and the off-resonant driving amplitudes ramp down with the intensity, leading to a much more robust transfer to the target state.

The total expected infidelity incurred by spin dephasing, spontaneous Raman scattering, and population of the intermediate states can thus be reduced below the $10^{-4}$ level at reasonable $\pi$ times. In principle, there does not appear to be anything fundamental preventing four- and six-photon single qubit gates from achieving fidelities demonstrated in state-of-the-art two-photon gates
~\cite{Gaebler2016,ballance2016}. Additionally, the intensity sensitivity of four-photon transitions should be of the same order as Raman-driven M\o lmer-S\o rensen entangling gates, so we expect the intensity stabilization requirements to be equivalent to previously demonstrated capabilities~\cite{Gaebler2016}. We note that the use of two non-copropagating beams in this work was to maximize the available intensity and is not strictly necessary.

Higher-order transitions naturally require more power than their two-photon counterparts, but we were able to achieve $\pi$ times below 100\,\textmu s with a large detuning of 44\,THz and beam waists of $\sim30$\,\textmu m. They can be made orders of magnitude faster by decreasing the beam waists, which is already required when moving to individually-addressed qubit systems, and $\Delta$. This does not come with the usual penalty of increasing spontaneous scattering probability per gate due to the stronger $\Delta$ scaling in Eqs.~\eqref{4photon_rabi} and ~\eqref{6photon_rabi}. In addition, at the near infrared wavelengths relevant for metastable qubit Raman, it is generally easier to produce high intensity laser beams and trap charging effects are much weaker. Finally, we expect the super-linear dependence of the Rabi frequencies on the beam intensities to lead to significant speed enhancements when driving these transitions with pulsed lasers.

\section*{Conclusion}\label{sec4}
These transitions can provide arbitrary unitary control between qu$d$it states with a clear path to fidelities in excess of $99.99\%$, which will enable more efficient qu$d$it control, state tomography, and simplified direct unitaries in single-atom quantum error correction codes like the absorption-emission code~\cite{AEcode} and spin-cat logical encodings~\cite{spincats, debry2026error}. Other applications may include simplified state preparation of large-spin atoms~\cite{Gaebler2016,ballance2016} and metrology with large spin superpositions~\cite{yang2025minute}. Additionally, the $(P_\perp P_\parallel)^{n/4}$ scaling of $n$-photon transition Rabi frequencies can be exploited to reduce cross-talk in individually addressed atom arrays~\cite{Hansch2001}. With growing interest in and emerging demonstrations of higher-dimensional quantum information processing, we believe that the implementation of high-fidelity four- and six-photon transitions may provide a useful tool towards achieving fault-tolerant quantum computing with a lower hardware overhead.

\backmatter

\bmhead{Acknowledgements}

We acknowledge useful discussions with K. Barajas, W. Campbell, E. Hudson, K. DeBry, I. Chuang, J. Chiaverini, V. Buchemmavari, and H. Haeffner.  This research is supported in part by the NSF through the Q-SEnSE Quantum Leap Challenge Institute, Award \#2016244 and the US Army Research Office under award W911NF-24-1-0379. The data supporting the figures in this article are available upon reasonable request from J.O.

\section*{Methods}
\subsection*{Derivation of higher order Rabi frequencies}
\label{sec:methods_derivation}
In this section we present a derivation of the four-photon Rabi frequency discussed in the main text by uncoupling the state amplitude equations of motion generated by the Schrodinger equation and taking an adiabatic elimination approximation. We also present heuristic expressions that can be used to find four- and six-photon Rabi frequencies of other systems.

The coupling of states $\ket{i}$ and $\ket{j}$ via the $k$'th beam with frequency $\omega_k$ is described by the $ij$'th component of the Hamiltonian
\begin{align}
    V_{i,j}(t) = -\frac{\hbar}{2}\Omega_{ij}e^{-i\omega_kt}
    \label{couping}
\end{align}

\noindent where $\Omega_{ij}$ is the complex single-beam Rabi frequency coupling the two states. In our treatment we include only terms of the laser-atom interaction Hamiltonian that correspond to the resonant dynamics, and we assume that off-resonant terms may be eliminated at the end of the calculation in a rotating wave approximation (RWA). With the full Hamiltonian $H(t)$ being the sum of the bare energies of the system $H_0$ and the interaction $V(t)$, we apply the Schr\"odinger equation and generate the following set of coupled differential equations for the state amplitudes $c_i$ with state labels defined in Fig.~\ref{fig:paths_flopping}:

\begin{align*}
    \dot{c_0'}&=c''_6\frac{i\Omega^*_{06}}{2}\\ 
    \dot{c_1''} +i\delta_1c_1''&=c''_6\frac{i\Omega^*_{16}}{2} + c''_7\frac{i\Omega^*_{17}}{2} \\
    \dot{c_2''} - i\delta_2c_2''&=c''_6\frac{i\Omega^*_{26}}{2} + c''_8\frac{i\Omega^*_{28}}{2} \\
    \dot{c_3'}&=c''_7\frac{i\Omega^*_{37}}{2} + c''_8\frac{i\Omega^*_{38}}{2}  \\
    \dot{c_6''} + i\Delta_6c_6''&=c'_0\frac{i\Omega_{06}}{2}  + c''_1\frac{i\Omega_{16}}{2} + c''_2\frac{i\Omega_{26}}{2}\\
    \dot{c_7''} + i\Delta_7c_7''&=c''_1\frac{i\Omega_{17}}{2}  + c'_3\frac{i\Omega_{37}}{2} \\
    \dot{c_8''} + i\Delta_8c_8''&=c''_2\frac{i\Omega_{28}}{2}  + c'_3\frac{i\Omega_{38}}{2}. \\
\end{align*}

We have rewritten the beam frequencies $\omega_r$ and $\omega_b$ in the coefficient for $\Omega_{ij}$ in terms of the state separation $\omega_{ij}$ and relevant beam detunings. For example, in the differential equation for $\dot{c_0}$, we can replace $\omega_r$ with $\omega_{06}-\Delta_6$ and in the equation for $\dot{c_2}$, we will replace $\omega_b$ with $\omega_{26}-\Delta_6 - \delta_2$. Additionally we have defined $\delta_1$ and $\delta_2$ as the magnitude of the detunings $\omega-\omega_{01}$ and $\omega_{01} + \omega_{12}-\omega$, respectively, with $\omega$ defined as the detuning of the two beams $\omega_b-\omega_r$. We have also transformed the two primary amplitudes of interest ($c_0$ and $c_3$) to the interaction representation, $c'_i = c_ie^{\frac{iE_it}{\hbar}}$. For the other states, we transformed to the field-interaction representation, the frame in which the state rotates at the detuning of the drive fields. The excited states in the $\mathrm{P_{3/2}}$ undergo the transformation $c''_i = c_ie^{i(\frac{E_i}{\hbar}-\Delta_i)t}$, and the intermediate states in the $\mathrm{D_{5/2}}$ transform to $c''_1 = c_1e^{i(\frac{E_1}{\hbar}-\delta_1)t}$ and $c''_2 = c_2e^{i(\frac{E_2}{\hbar}+\delta_2)t}$.

Next we perform the adiabatic elimination~\cite{daveRaman}  of the $\mathrm{P_{3/2}}$ states and intermediate states in the $\mathrm{D_{5/2}}$ manifold by noting that the time derivatives in the equations for $i=1,2,6,7,8$ are small compared to the $\mathrm{P_{3/2}}$ state detunings $\Delta$ and the intermediate $\mathrm{D_{5/2}}$ state detunings $\delta$, so our equations of motion reduce to
\begin{align*}
    \dot{c_0'}&=c''_6\frac{i\Omega^*_{06}}{2}\\ 
    c''_1 &\simeq \frac{1}{2\delta_1} \left(\Omega^*_{16}c''_6 + \Omega^*_{17}c''_7 \right)\\
    c''_2 &\simeq -\frac{1}{2\delta_2} \left(\Omega^*_{26}c''_6 + \Omega^*_{28}c''_8 \right)\\
    \dot{c_3'}&=c''_7\frac{i\Omega^*_{37}}{2} + c''_8\frac{i\Omega^*_{38}}{2}\\
    c''_6 &\simeq \frac{1}{2\Delta_6}\left(c'_0\Omega_{06}  + c''_1\Omega_{16} + c''_2\Omega_{26}\right)\\
    c''_7 &\simeq \frac{1}{2\Delta_7}\left(c''_1\Omega_{17}  + c'_3\Omega_{37}\right)\\
    c''_8 &\simeq \frac{1}{2\Delta_8}\left(c''_2\Omega_{28}  + c'_3\Omega_{38}\right).\\
\end{align*}
Finally, we can uncouple the system of linear equations to reduce the problem to a single differential equation for amplitudes $c'_0$ and $c'_3$ of the form
\begin{align}
    \dot{c'_0} = iCc'_0 + i\frac{\mathbf{\Omega_{03}^{(4)}}}{2}c'_3
\end{align}
\noindent where $C$ is a term involving a.c. Stark shifts, and
\begin{align}
\label{four_photon_methods}
   \boxed{
   \mathbf{\Omega_{03}^{(4)}}=\frac{\Omega^*_{06}}{8\Delta^2}\left(\frac{\Omega_{16}\Omega^*_{17}\Omega_{37}}{\delta_1}-\frac{\Omega_{26}\Omega^*_{28}\Omega_{38}}{\delta_2}\right).}
\end{align}

Taking note of the pattern we write out a heuristic expression for generating the n-photon Rabi frequency between states $a$ and $b$.:
\begin{align}
    \Omega^{(2)}_{ab}&=\sum_{i}{\frac{\Omega_{ai}\Omega_{ib}}{2\delta_i}}\\
        \Omega^{(4)}_{ab}&=\sum_{i, j, k}{\frac{\Omega_{ai}\Omega_{ij}\Omega_{jk}\Omega_{kb}}{2^3\delta_i\delta_j\delta_k}}\\
        \Omega^{(6)}_{ab}&=\sum_{i, j, k, l, m}{\frac{\Omega_{ai}\Omega_{ij}\Omega_{jk}\Omega_{kl}\Omega_{lm}\Omega_{mb}}{2^5\delta_i\delta_j\delta_k\delta_l\delta_m}}\label{six_photon_general}
\end{align}
We use Equation~\eqref{six_photon_general} to generate Equation~\eqref{6photon_rabi} in the main text. A more complete derivation of~\ref{four_photon_methods} is presented in Appendix~\ref{appendix:fourphotonderivation}.

\subsection*{State preparation and measurement}
\label{sec:methods_spam}
\begin{figure}[h]
\centering$\mathrm{P_{3/2}}$
    \hspace*{-0.0in}
    \includegraphics[width = \columnwidth]{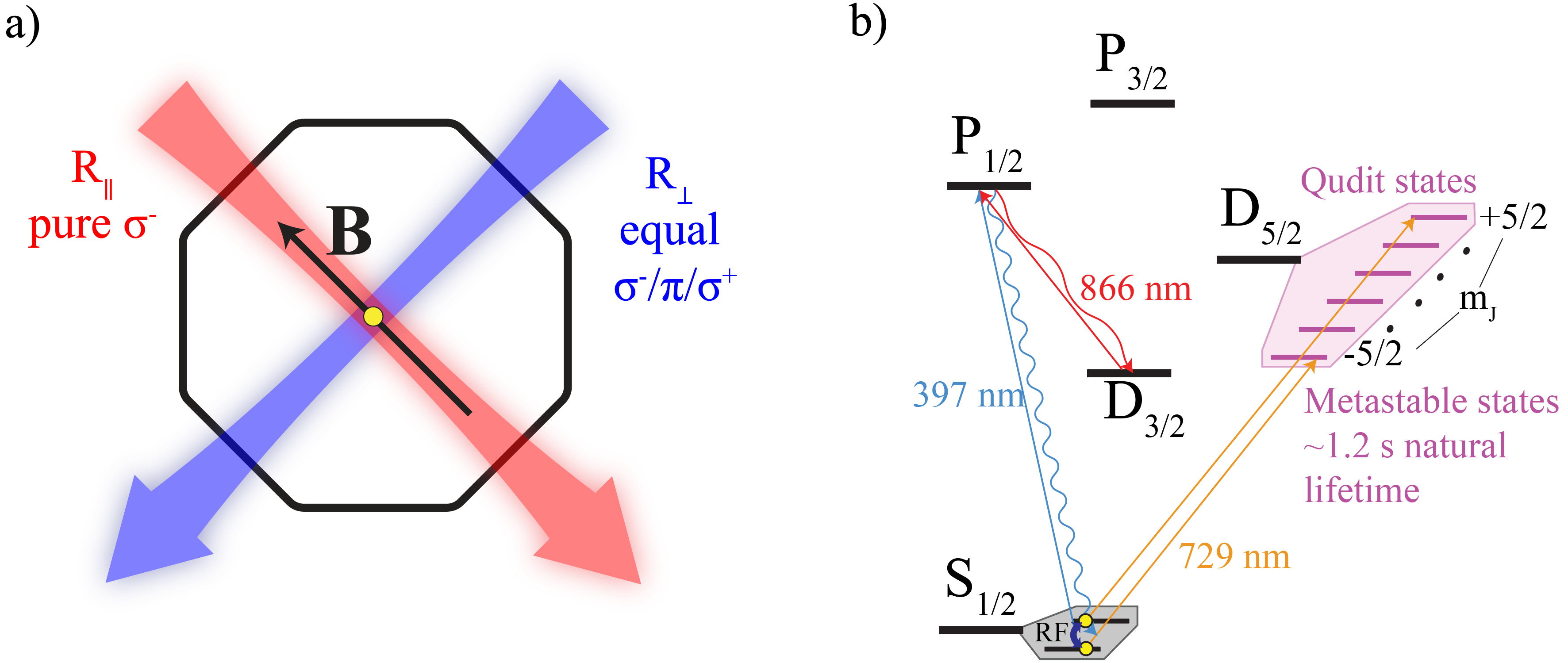}
    \caption{\textbf{a)} Geometry of 976\,nm Raman beams used to drive four- and six-photon transitions relative to quantization magnetic field direction. \textbf{b)} Relevant atomic structure of $^{40}$Ca$^+$ used in this work. 397\,nm beam drives fluorescent $\mathrm{S_{1/2}}\leftrightarrow \mathrm{P_{1/2}}$ cycling transition with 866\,nm beam to depump the $\mathrm{D_{3/2}}$ manifold. 729\,nm beam drives $\pi$ pulses between $\mathrm{S}$ and $\mathrm{m_J}$ qudit states encoded in $\mathrm{D_{5/2}}$ for state-preparation and readout of the qudit states. An optional RF $\pi$ pulse between $\mathrm{S_{1/2}}$ ground states is used to maximize the 729 transition Rabi frequency to each qudit state.}
\label{fig:geometry}
\end{figure}
To prepare the ion in an arbitrary sublevel of the D$_{5/2}$ manifold, we first Doppler and electromagnetically-induced transparency (EIT) cool~\cite{EIT} its motion to $\overline{n}<1$ and then optically pump it to the $\ket{S_{1/2},\mathrm{m_J}=+1/2}$ state using 397\,nm and 866\,nm laser beams. Then, we drive a narrow, state-selective, 729\,nm $\mathrm{S_{1/2}}\leftrightarrow \mathrm{D_{5/2}}$ electric quadrupole transition preceded by an optional rf $\pi$ pulse to $\ket{\mathrm{S_{1/2}},\mathrm{m_J}=-1/2}$. The initial $S_{1/2}$ sublevel is chosen to maximize the Rabi frequency. After this shelving process~\cite{shelving}, we perform a fluorescence check for 500\,\textmu s and repeat the process until the ion is dark, indicating successful preparation in $\mathrm{D_{5/2}}$~\cite{Sotirova2024}. 

We read out each sublevel of the $\mathrm{D_{5/2}}$ manifold by deshelving them in turn to the $\mathrm{S_{1/2}}$ manifold and checking for fluorescence on the 397\,nm cycling transition with the $866$\,nm repumping beam on. The first state to produce a bright signal is considered the measurement result, but imperfect deshelving pulses can lead to the ion remaining dark through all the checks. These shots are discarded and lead to a qu$d$it readout inefficiency of $\sim0.5\%$. For the remaining trials we achieve state preparation and measurement fidelities above 99.5\% for all states within the $\mathrm{D_{5/2}}$ manifold. 

\begin{appendices}

\setcounter{section}{0}    
\renewcommand{\thesection}{\arabic{section}}




\section{Four-photon Rabi frequency derivation}
\label{appendix:fourphotonderivation}
We present two methods to derive the Rabi frequencies of the higher order four- and six-photon transitions and provide example derivations for the exemplary four-photon transition in the main text. The first method follows the standard treatment of the two-photon Raman Rabi frequency derivation, wherein state amplitude equations of motion are generated via the Schr\"odinger equation, an adiabatic elimination approximation is taken to eliminate the virtual excited and intermediate states, and the equations of motion of the effective two-state system can be uncoupled. This method, however, can present ambiguities in the choice of frame to rotate to. Although not present in the example four-photon derivation given here, this does appear in the six-photon process we investigated. The second method employs time-dependent perturbation theory to calculate the leading order corrections to the time evolution of the state amplitudes and does not suffer the ambiguity sometimes present in the first method. Both methods arrive at the same result.

\vspace{2em}
\textbf{\textit{Differential equation approach}}

The coupling of states $\ket{i}$ and $\ket{j}$ via the $k$'th beam with frequency $\omega_k$ is described by the $ij$'th component of the Hamiltonian
\begin{align}
    V_{i,j} = -\frac{\hbar}{2}\Omega_{ij}e^{i\omega_kt}
    \label{coupling}
\end{align}
where $\Omega_{ij}$ is the complex single-beam Rabi frequency coupling the two states. In our treatment we include only terms of the laser-atom interaction Hamiltonian that correspond to the resonant dynamics, and we assume that off-resonant terms may be eliminated at the end of the calculation in a rotating wave approximation (RWA). We can then write out the full interaction Hamiltonian of the 7-level system shown in Fig.~\ref{fig:paths_flopping}(a) in matrix form:

\begin{align}
    \label{matrices}
    \renewcommand\arraystretch{2}
    V(t)=-\frac{\hbar}{2}\begin{pmatrix}
            0 & 0 & 0 & 0 & \Omega^*_{06}e^{i\omega_rt} & 0 & 0\\
            0 & 0 & 0 & 0 & \Omega^*_{16}e^{i\omega_bt} & \Omega^*_{17}e^{i\omega_rt} & 0\\
            0 & 0 & 0 & 0 & \Omega^*_{26}e^{i\omega_bt} & 0 & \Omega^*_{28}e^{i\omega_rt}\\
            0 & 0 & 0 & 0 & 0 & \Omega^*_{37}e^{i\omega_bt} & \Omega^*_{38}e^{i\omega_bt}\\
            \Omega_{06}e^{-i\omega_rt} & \Omega_{16}e^{-i\omega_bt} & \Omega_{26}e^{-i\omega_bt} & 0 & 0 & 0 & 0\\
            0 & \Omega_{17}e^{-i\omega_rt} & 0  & \Omega_{37}e^{-i\omega_bt} & 0 & 0 & 0\\
            0 & 0 & \Omega_{28}e^{-i\omega_rt} & \Omega_{38}e^{-i\omega_bt} & 0 & 0 & 0
        \end{pmatrix}.
\end{align}
\newline

\noindent where the state labels are defined in Fig.~\ref{fig:paths_flopping}. By applying the Schr\"odinger equation with time evolution governed by Equation~\eqref{matrices}, we generate the following set of coupled differential equations for the state amplitudes $c_i$:
    
\begin{align*}
    \dot{c_0}&=c_6\frac{i\Omega^*_{06}e^{i\omega_rt}}{2}-\frac{iE_0}{\hbar}c_0 \\ 
    \dot{c_1}&=c_6\frac{i\Omega^*_{16}e^{i\omega_bt}}{2} + c_7\frac{i\Omega^*_{17}e^{i\omega_rt}}{2} -\frac{iE_1}{\hbar}c_1 \\
    \dot{c_2}&=c_6\frac{i\Omega^*_{26}e^{i\omega_bt}}{2} + c_8\frac{i\Omega^*_{28}e^{i\omega_rt}}{2} -\frac{iE_2}{\hbar}c_2 \\
    \dot{c_3}&=c_7\frac{i\Omega^*_{37}e^{i\omega_bt}}{2} + c_8\frac{i\Omega^*_{38}e^{i\omega_bt}}{2} -\frac{iE_3}{\hbar}c_3 \\
    \dot{c_6}&=c_0\frac{i\Omega_{06}e^{-i\omega_rt}}{2} + c_1\frac{i\Omega_{16}e^{-i\omega_bt}}{2} + c_2\frac{i\Omega_{26}e^{-i\omega_bt}}{2} -\frac{iE_6}{\hbar}c_6 \\
    \dot{c_7}&=c_1\frac{i\Omega_{17}e^{-i\omega_rt}}{2} + c_3\frac{i\Omega_{37}e^{-i\omega_bt}}{2} -\frac{iE_7}{\hbar}c_7 \\
    \dot{c_8}&=c_2\frac{i\Omega_{28}e^{-i\omega_rt}}{2} + c_3\frac{i\Omega_{38}e^{-i\omega_bt}}{2} -\frac{iE_8}{\hbar}c_8 .\\
\end{align*}

Anticipating a picture change, we rewrite the beam frequencies $\omega_r$ and $\omega_b$ in the coefficient for $\Omega_{ij}$ in terms of the state separation $\omega_{ij}$ and relevant beam detunings. For example, in the differential equation for $\dot{c_0}$, we can replace $\omega_r$ with $\omega_{06}-\Delta_6$ and in the equation for $\dot{c_2}$, we will replace $\omega_b$ with $\omega_{26}-\Delta_6 - \delta_2$. We can then re-write the coupled differential equations as

\begin{align*}
    \dot{c_0}&=c_6\frac{i\Omega^*_{06}e^{i(\omega_{06}-\Delta_6)t}}{2}-\frac{iE_0}{\hbar}c_0 \\ 
    \dot{c_1}&=c_6\frac{i\Omega^*_{16}e^{i(\omega_{16}-\Delta_6 + \delta_1)t}}{2} + c_7\frac{i\Omega^*_{17}e^{i(\omega_{17}-\Delta_7 + \delta_1)t}}{2} -\frac{iE_1}{\hbar}c_1 \\
    \dot{c_2}&=c_6\frac{i\Omega^*_{26}e^{i(\omega_{26}-\Delta_6 - \delta_2)t}}{2} + c_8\frac{i\Omega^*_{28}e^{i(\omega_{28}-\Delta_8 - \delta_2)t}}{2} -\frac{iE_2}{\hbar}c_2 \\
    \dot{c_3}&=c_7\frac{i\Omega^*_{37}e^{i(\omega_{37}-\Delta_7)t}}{2} + c_8\frac{i\Omega^*_{38}e^{i(\omega_{38}-\Delta_8)t}}{2} -\frac{iE_3}{\hbar}c_3 \\
    \dot{c_6}&=c_0\frac{i\Omega_{06}e^{-i(\omega_{06}-\Delta_6)t}}{2} + c_1\frac{i\Omega_{16}e^{-i(\omega_{16}-\Delta_6 +\delta_1)t}}{2} + c_2\frac{i\Omega_{26}e^{-i(\omega_{26}-\Delta_6 - \delta_2)t}}{2} -\frac{iE_6}{\hbar}c_6 \\
    \dot{c_7}&=c_1\frac{i\Omega_{17}e^{-i(\omega_{17}-\Delta_7 + \delta_1)t}}{2} + c_3\frac{i\Omega_{37}e^{-i(\omega_{37}-\Delta_7)t}}{2} -\frac{iE_7}{\hbar}c_7 \\
    \dot{c_8}&=c_2\frac{i\Omega_{28}e^{-i(\omega_{28}-\Delta_8 - \delta_2)t}}{2} + c_3\frac{i\Omega_{38}e^{-i(\omega_{38}-\Delta_8)t}}{2} -\frac{iE_8}{\hbar}c_8 \\
\end{align*}
where we have defined $\delta_1$ and $\delta_2$ as the magnitude of the detunings $\omega-\omega_{01}$ and $\omega_{01} + \omega_{12}-\omega$, respectively, with $\omega$ defined as the detuning of the two beams $\omega_b-\omega_r$. 

We now look to write the state amplitudes in a more illuminating representation. The mathematics simplifies if we transform the two primary amplitudes of interest ($c_0$ and $c_3$) to the interaction representation, $c'_i = c_ie^{\frac{iE_it}{\hbar}}$. For the other states, we will transform to the field-interaction representation, the frame in which the state rotates at the detuning of the drive fields. The excited states in the $\mathrm{P_{3/2}}$ will transform to $c''_i = c_ie^{i(\frac{E_i}{\hbar}-\Delta_i)t}$, and the intermediate states in the $\mathrm{D_{5/2}}$ will transform to $c''_1 = c_1e^{i(\frac{E_1}{\hbar}-\delta_1)t}$ and $c''_2 = c_2e^{i(\frac{E_2}{\hbar}+\delta_2)t}$. The coupled equations of motion are then

\begin{align*}
    \dot{c_0'}&=c''_6\frac{i\Omega^*_{06}}{2}\\ 
    \dot{c_1''} +i\delta_1c_1''&=c''_6\frac{i\Omega^*_{16}}{2} + c''_7\frac{i\Omega^*_{17}}{2} \\
    \dot{c_2''} - i\delta_2c_2''&=c''_6\frac{i\Omega^*_{26}}{2} + c''_8\frac{i\Omega^*_{28}}{2} \\
    \dot{c_3'}&=c''_7\frac{i\Omega^*_{37}}{2} + c''_8\frac{i\Omega^*_{38}}{2}  \\
    \dot{c_6''} + i\Delta_6c_6''&=c'_0\frac{i\Omega_{06}}{2}  + c''_1\frac{i\Omega_{16}}{2} + c''_2\frac{i\Omega_{26}}{2}\\
    \dot{c_7''} + i\Delta_7c_7''&=c''_1\frac{i\Omega_{17}}{2}  + c'_3\frac{i\Omega_{37}}{2} \\
    \dot{c_8''} + i\Delta_8c_8''&=c''_2\frac{i\Omega_{28}}{2}  + c'_3\frac{i\Omega_{38}}{2}. \\
\end{align*}
\newline
We adiabatically eliminate~\cite{daveRaman} the $\mathrm{P_{3/2}}$ states and intermediate states in the $\mathrm{D_{5/2}}$ manifold by noting that the time derivatives in the equations for $i=1,2,6,7,8$ are small compared to the $\mathrm{P_{3/2}}$ state detunings $\Delta$ and the intermediate $\mathrm{D_{5/2}}$ state detunings $\delta$, so our equations of motion reduce to

\begin{align*}
    \dot{c_0'}&=c''_6\frac{i\Omega^*_{06}}{2}\\ 
    c''_1 &\simeq \frac{1}{2\delta_1} \left(\Omega^*_{16}c''_6 + \Omega^*_{17}c''_7 \right)\\
    c''_2 &\simeq -\frac{1}{2\delta_2} \left(\Omega^*_{26}c''_6 + \Omega^*_{28}c''_8 \right)\\
    \dot{c_3'}&=c''_7\frac{i\Omega^*_{37}}{2} + c''_8\frac{i\Omega^*_{38}}{2}\\
    c''_6 &\simeq \frac{1}{2\Delta_6}\left(c'_0\Omega_{06}  + c''_1\Omega_{16} + c''_2\Omega_{26}\right)\\
    c''_7 &\simeq \frac{1}{2\Delta_7}\left(c''_1\Omega_{17}  + c'_3\Omega_{37}\right)\\
    c''_8 &\simeq \frac{1}{2\Delta_8}\left(c''_2\Omega_{28}  + c'_3\Omega_{38}\right).\\
\end{align*}

Finally, we can uncouple the system of linear equations to reduce the problem to a single differential equation for amplitudes $c'_0$ and $c'_3$ of the form
\begin{align}
    \dot{c'_0} = iCc'_0 + i\frac{\mathbf{\Omega_{03}^{(4)}}}{2}c'_3
\end{align}
    
\noindent where $C$ is a term involving a.c. Stark shifts, and

\begin{align}
\label{four_photon}
   \boxed{
   \mathbf{\Omega_{03}^{(4)}}=\frac{\Omega^*_{06}}{8\Delta^2}\left(\frac{\Omega_{16}\Omega^*_{17}\Omega_{37}}{\delta_1}-\frac{\Omega_{26}\Omega^*_{28}\Omega_{38}}{\delta_2}\right)}
\end{align}
\noindent is the four-photon Rabi frequency. The derivation for the six-photon Rabi frequency follows analogously.

\vspace{2em}
\textbf{\textit{Time-dependent perturbation theory}}
\newline
\newline

In this section we present a derivation of Equation~\eqref{four_photon} via time dependent perturbation theory. We will work in the interaction representation. The Hamiltonian of the system is
\begin{align}
    H = H_0 + V(t)
\end{align}
where $V(t)$ is defined in Equation~\eqref{matrices} and $H_0$ is the bare Hamiltonian of the ion. We first transform $V(t)$ into the interaction representation:

\begin{align}
    V_I(t) \equiv e^{iH_0t/\hbar}V(t)e^{-iH_0t/\hbar}
\end{align}
which, after substituting in $\delta_{0/1}$ as defined in the previous section, becomes
\[\scalebox{0.75}{$
V_I(t) = \frac{\hbar}{2}
\begin{pmatrix}
0 & 0 & 0 & 0 
  & \Omega_{06r} e^{-i\Delta t} 
  & 0 & 0 \\[6pt]

0 & 0 & 0 & 0 
  & \Omega_{16b} e^{-i(\Delta - \delta_1)t} 
  & \Omega_{17r} e^{-i(\Delta - \delta_1)t} 
  & 0 \\[6pt]

0 & 0 & 0 & 0 
  & \Omega_{26b} e^{-i(\Delta + \delta_2)t} 
  & 0 
  & \Omega_{28r} e^{-i(\Delta + \delta_2)t} \\[6pt]

0 & 0 & 0 & 0 
  & 0 
  & \Omega_{37b} e^{-i\Delta t} 
  & \Omega_{38b} e^{-i\Delta t} \\[6pt]

\Omega_{06r} e^{i\Delta t} 
  & \Omega_{16b} e^{i(\Delta - \delta_1)t} 
  & \Omega_{26b} e^{i(\Delta + \delta_2)t} 
  & 0 & 0 & 0 & 0 \\[6pt]

0 
  & \Omega_{17r} e^{i(\Delta - \delta_1)t} 
  & 0 
  & \Omega_{37b} e^{i\Delta t} 
  & 0 & 0 & 0 \\[6pt]

0 & 0 
  & \Omega_{28r} e^{i(\Delta + \delta_2)t} 
  & \Omega_{38b} e^{i\Delta t} 
  & 0 & 0 & 0
\end{pmatrix}.$}
\]

Now that we have the interaction Hamiltonian, we can solve for the time evolution propagator $U(t)$. If we initialize the state in $\ket{0}$ at time $t = 0$, $U(t)$ is defined as
\begin{align}
    \ket{\psi(t)} = U(t)\ket{0}
\end{align}
and the probability of the state being measured in state $\ket{3}$ after some time $t$ is
\begin{align}
    P_{0\rightarrow3}(t) = \left|\braket{3|\psi(t)}\right|^2 = \left|\braket{3|U(t)|0}\right|^2.
\end{align}

To obtain an expression for the matrix element $\braket{3|U(t)|0}$, we first expand $U(t)$ in a Dyson series:
\begin{align}
    U(t) = U^{(0)}(t) + U^{(1)}(t) + U^{(2)}(t) + U^{(3)}(t) + U^{(4)}(t) + \dots
\end{align}
with
\begin{align*}
    U^{(0)}(t) &= 1 \\
    U^{(1)}(t) &= \frac{-i}{\hbar}\int_0^tdt'V_I(t')\\
    U^{(2)}(t) &= \left(\frac{-i}{\hbar}\right)^2\int_0^tdt'\int_0^{t'}dt''V_I(t')V_I(t'')\\
    U^{(3)}(t) &= \left(\frac{-i}{\hbar}\right)^3\int_0^tdt'\int_0^{t'}dt''\int_0^{t''}dt'''V_I(t')V_I(t'')V_I(t''')\\
    U^{(4)}(t) &= \left(\frac{-i}{\hbar}\right)^4\int_0^tdt'\int_0^{t'}dt''\int_0^{t''}dt'''\int_0^{t'''}dt''''V_I(t')V_I(t'')V_I(t''')V_I(t'''')\\
    \vdots
\end{align*}

The matrix element $\braket{3|U^{(i)}(t)|0}$ vanishes for all $i<4$, a consequence of the angular momentum constraint of the dipole-coupled photons, and the leading correction to the propagator is therefore
\begin{align*}
    \bra{3}U(t)\ket{0}\approx&\frac{\Omega_{06r}\,\Omega_{16b}\,\Omega_{17r}\,\Omega_{37b}}{16}\\
    \times&\int_0^tdt'\int_0^{t'}dt''\int_0^{t''}dt'''\int_0^{t'''}dt''''e^{-i t' \Delta + i t'''' \Delta 
    + i t''(\Delta - \delta_1)
    - i t'''(\Delta - \delta_1)}\\
    +&\frac{\Omega_{06r}\,\Omega_{26b}\,\Omega_{28r}\,\Omega_{38b}}{16}\\
    \times&\int_0^tdt'\int_0^{t'}dt''\int_0^{t''}dt'''\int_0^{t'''}dt''''e^{-i t' \Delta + i t'''' \Delta 
    + i t''(\Delta + \delta_{2})
    - i t'''(\Delta + \delta_{2})}
\end{align*}

Working through these integrals we find 

\begin{align*}
\bra{3}U(t)\ket{0}\approx&\textcolor{myorange}{\frac{3 \,\Omega_{06r} \Omega_{16b} \Omega_{17r} \Omega_{37b}}
      {16 \,\Delta^2 (\Delta - \delta_1)^2}}
-
\textcolor{myorange}{\frac{3 e^{-i t \Delta} \,\Omega_{06r} \Omega_{16b} \Omega_{17r} \Omega_{37b}}
      {16 \,\Delta^2 (\Delta - \delta_1)^2}}
-
\textcolor{myorange}{\frac{i t \,\Omega_{06r} \Omega_{16b} \Omega_{17r} \Omega_{37b}}
      {8 \,\Delta (\Delta - \delta_1)^2}} \\[1mm]
&-
\textcolor{myorange}{\frac{i e^{-i t \Delta} t \,\Omega_{06r} \Omega_{16b} \Omega_{17r} \Omega_{37b}}
      {16 \,\Delta (\Delta - \delta_1)^2}}
-
\textcolor{myblue}{\frac{\,\Omega_{06r} \Omega_{16b} \Omega_{17r} \Omega_{37b}}
      {16 (\Delta - \delta_1)^2 \delta_1^2}} \\[1mm]
&+
\textcolor{myblue}{\frac{e^{-i t \delta_1} \,\Omega_{06r} \Omega_{16b} \Omega_{17r} \Omega_{37b}}
      {16 (\Delta - \delta_1)^2 \delta_1^2}} 
+
\fcolorbox{black}{white}{$\displaystyle\frac{i t \,\Omega_{06r} \Omega_{16b} \Omega_{17r} \Omega_{37b}}
      {16 (\Delta - \delta_1)^2 \delta_1} $} \\[1mm]
&-
\textcolor{myorange}{\frac{\delta_1 \,\Omega_{06r} \Omega_{16b} \Omega_{17r} \Omega_{37b}}
      {8 \,\Delta^3 (\Delta - \delta_1)^2} }
+
\textcolor{myorange}{\frac{e^{-i t \Delta} \delta_1 \,\Omega_{06r} \Omega_{16b} \Omega_{17r} \Omega_{37b}}
      {8 \,\Delta^3 (\Delta - \delta_1)^2}} \\[1mm]
&+
\textcolor{myorange}{\frac{i t \delta_1 \,\Omega_{06r} \Omega_{16b} \Omega_{17r} \Omega_{37b}}
      {16 \,\Delta^2 (\Delta - \delta_1)^2} }
+
\textcolor{myorange}{\frac{i e^{-i t \Delta} t \delta_1 \,\Omega_{06r} \Omega_{16b} \Omega_{17r} \Omega_{37b}}
      {16 \,\Delta^2 (\Delta - \delta_1)^2}} \\[2mm]
&+
\textcolor{myorange}{\frac{3 \,\Omega_{06r} \Omega_{26b} \Omega_{28r} \Omega_{38b}}
      {16 \,\Delta^2 (\Delta + \delta_2)^2}}
-
\textcolor{myorange}{\frac{3 e^{-i t \Delta} \,\Omega_{06r} \Omega_{26b} \Omega_{28r} \Omega_{38b}}
      {16 \,\Delta^2 (\Delta + \delta_2)^2}} 
-
\textcolor{myorange}{\frac{i t \,\Omega_{06r} \Omega_{26b} \Omega_{28r} \Omega_{38b}}
      {8 \,\Delta (\Delta + \delta_2)^2}} \\[1mm]
&-
\textcolor{myorange}{\frac{i e^{-i t \Delta} t \,\Omega_{06r} \Omega_{26b} \Omega_{28r} \Omega_{38b}}
      {16 \,\Delta (\Delta + \delta_2)^2}}
-
\textcolor{mygreen}{\frac{\,\Omega_{06r} \Omega_{26b} \Omega_{28r} \Omega_{38b}}
      {16 \,\delta_2^2 (\Delta + \delta_2)^2} } \\[1mm]
&+
\textcolor{mygreen}{\frac{e^{-i t \Delta + i t (\Delta + \delta_2)} \,\Omega_{06r} \Omega_{26b} \Omega_{28r} \Omega_{38b}}
      {16 \,\delta_2^2 (\Delta + \delta_2)^2}} 
-
\fcolorbox{black}{white}{$\displaystyle\frac{i t \,\Omega_{06r} \Omega_{26b} \Omega_{28r} \Omega_{38b}}
      {16 \,\delta_2 (\Delta + \delta_2)^2}$} \\[1mm]
&+
\textcolor{myorange}{\frac{\delta_2 \,\Omega_{06r} \Omega_{26b} \Omega_{28r} \Omega_{38b}}
      {8 \,\Delta^3 (\Delta + \delta_2)^2} }
-
\textcolor{myorange}{\frac{e^{-i t \Delta} \delta_2 \,\Omega_{06r} \Omega_{26b} \Omega_{28r} \Omega_{38b}}
      {8 \,\Delta^3 (\Delta + \delta_2)^2}} \\[1mm]
&-
\textcolor{myorange}{\frac{i t \delta_2 \,\Omega_{06r} \Omega_{26b} \Omega_{28r} \Omega_{38b}}
      {16 \,\Delta^2 (\Delta + \delta_2)^2} }
-
\textcolor{myorange}{\frac{i e^{-i t \Delta} t \delta_2 \,\Omega_{06r} \Omega_{26b} \Omega_{28r} \Omega_{38b}}
      {16 \,\Delta^2 (\Delta + \delta_2)^2}}.
\end{align*}

\begin{figure}[h]
\centering
    \hspace*{-0.0in}
    \includegraphics[width = 0.6\columnwidth]{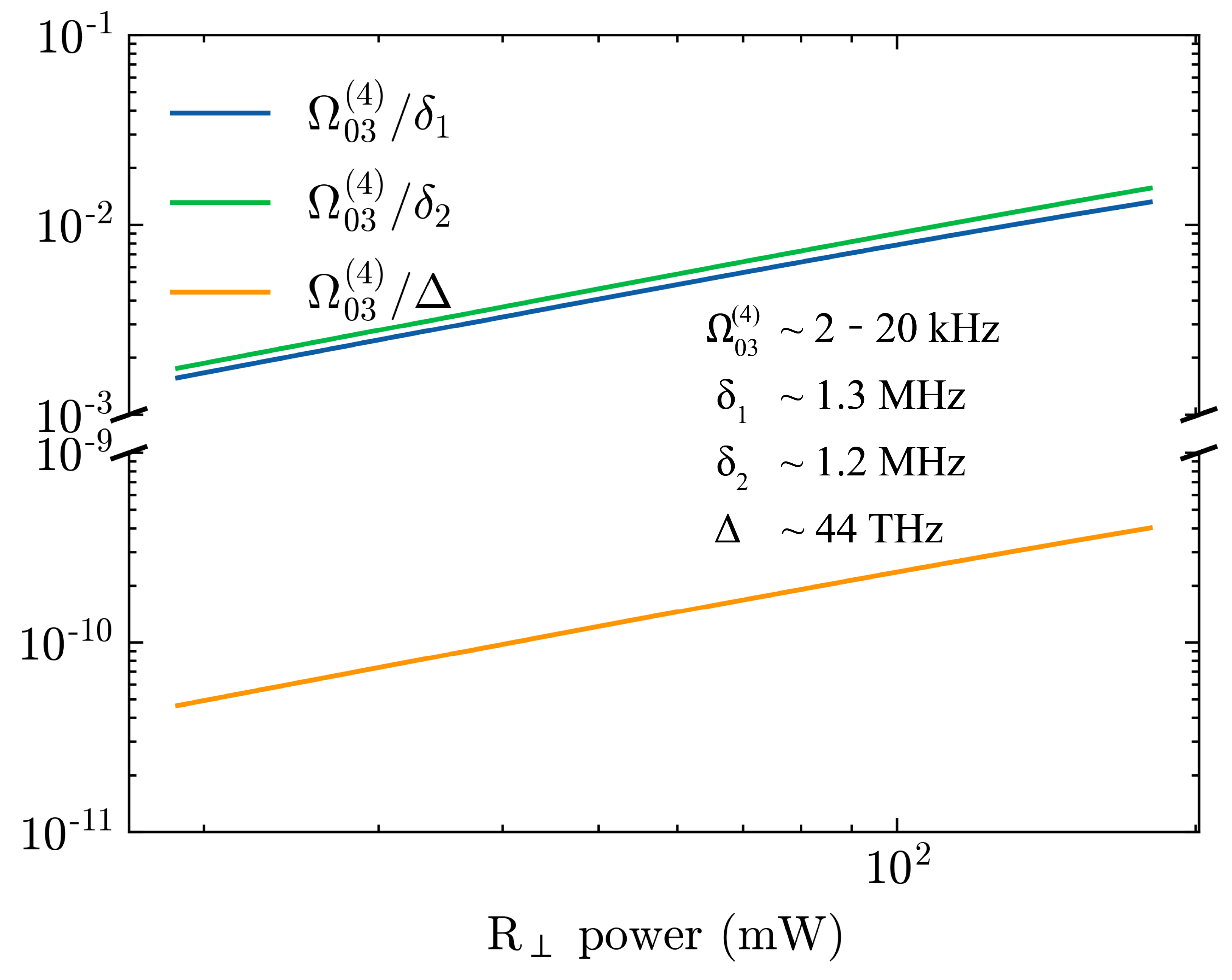}
    \caption{Ratio of four-photon Rabi frequency (analytic) to detuning from $\mathrm{P}$ and intermediate $\mathrm{D}$ states. Adiabatic elimination approximation for such states (and corresponding terms in $\bra{3}U(t)\ket{0}$) is valid in the regime that the curve is $\ll1$. Colored terms in the expression for $\bra{3}U(t)\ket{0}$ are approximated to be zero, justified by corresponding colored curve in the figure.}
\label{fig:adiabatic_approx}
\end{figure}

At this point, we will make some approximations to simplify the expression. In our system, the $\mathrm{P}$ state detuning is much greater than the detuning from the intermediate states: $\Delta \gg\delta$. In addition, we must still make an approximation analogous to the adiabatic elimination approximation: the Rabi frequency of the transition is much slower than the detuning from the states that off-resonantly facilitate the transition ($\mathrm{P}$ states and intermediate states in $\mathrm{D_{5/2}}$). The orange terms above are approximately zero in the regime where $\mathbf{\Omega_{03}^{(4)}}/\Delta\ll1$, and the blue and green terms can be approximated as zero in the regime where $\mathbf{\Omega_{03}^{(4)}}/\delta_{1/2}\ll1$. Thus, we have
\begin{align}
   \bra{3}U(t)\ket{0}\approx\frac{\Omega^*_{06}t}{16\Delta^2}\left(\frac{\Omega_{16}\Omega^*_{17}\Omega_{37}}{\delta_1}-\frac{\Omega_{26}\Omega^*_{28}\Omega_{38}}{\delta_2}\right)
\end{align}

and we can make the definition

\begin{align}
    \boxed{\mathbf{\Omega_{03}^{(4)}}\equiv\frac{\Omega^*_{06}}{8\Delta^2}\left(\frac{\Omega_{16}\Omega^*_{17}\Omega_{37}}{\delta_1}-\frac{\Omega_{26}\Omega^*_{28}\Omega_{38}}{\delta_2}\right)}\,\,.
\end{align}
The probability to transfer population from $\ket{0}$ to $\ket{3}$ for short times $t$ is 

\begin{align}
    P_{0\rightarrow3}(t) = \left|\braket{3|U(t)|0}\right|^2\approx\frac{\left|\mathbf{\Omega_{03}^{(4)}}t\right|^2}{4}.
\end{align}

To show that we have obtained the four-photon Rabi frequency, we take note of the Taylor expansion

\begin{align}
    \sin^2{\left(\frac{x}{2}\right)} = \frac{x^2}{4} - \frac{x^4}{48} + \dots
\end{align}

and we posit that higher orders of perturbation theory will generate the higher order terms in the Taylor expansion for $\sin^2{\frac{x}{2}}$. This implies that the probability to measure the state in $\ket{3}$ after preparing in $\ket{0}$ and driving for a time $t$ evolves as

\begin{align}
    P_{0\rightarrow3}(t) = \sin^2{\left(\frac{\mathbf{\Omega_{03}^{(4)}}t}{2}\right)}
\end{align}

with $\mathbf{\Omega_{03}^{(4)}}$ the same four-photon Rabi frequency we calculated in the previous section.


\section{Numerical simulations}
\label{appendix:numericalsims}
We seek to validate the adiabatic elimination of the intermediate $\mathrm{D_{5/2}}$ states, which is the new, non-standard approximation made in this work. To this end, we numerically simulate the dynamics under a Hamiltonian considering couplings between states within $\mathrm{D_{5/2}}$ and $\mathrm{P_{3/2}}$. We reduce the problem to a six-level system by adiabatically eliminating the $\mathrm{P}$ states but not the D states. In this Hamiltonian we do not drop any fast rotating terms, as we did in Equation~\eqref{matrices}, and include all possible two-photon couplings.

\begin{figure}[h]
\centering
    \hspace*{-0.0in}
    \includegraphics[width = 0.99\columnwidth]{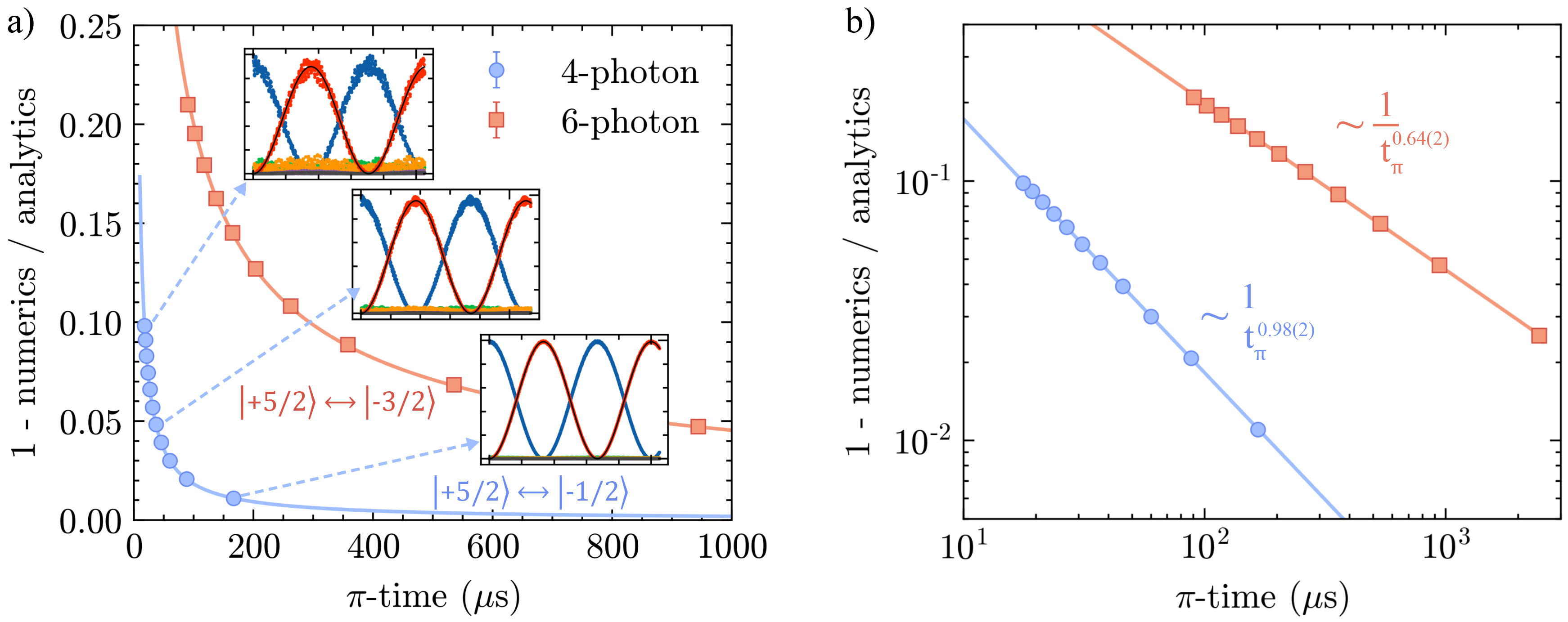}
    \caption{\textbf{a)} Numerically simulated four- (blue) and six- (red) photon Rabi frequencies divided by analytic predictions (Eqs.~\eqref{4photon_rabi}and ~\eqref{6photon_rabi}) at various driving strengths. As the transition $\pi$-time increases (decreasing beam power) the analytic and numeric predictions converge, justifying the $\mathrm{D_{5/2}}$ adiabatic elimination approximation. Inset figures are numerically-simulated resonant dynamics of the four-photon transition. \textbf{b)} Ratio of numerically simulated - analytic four- and six-photon Rabi frequencies on a log-log scale. A power law well describes the disagreement betwen numerics and analytics with increasing $\pi$-time, suggesting convergence between the analytic prediction and simulation in the limit of large $\pi$-time.}
\label{fig:convergence}
\end{figure}

We see a direct breakdown in the approximation at high beam powers, resulting in fast, small-amplitude dynamics on top of the main Rabi flopping envelope and non-negligible population of intermediate states (see Fig.~\ref{fig:intermediatestatepopulation}). In Fig.~\ref{fig:convergence}, we plot the ratio of the four-photon (six-photon) Rabi frequency as predicted by our numerical simulations to the result of Equation~\eqref{4photon_rabi} (Equation~\eqref{6photon_rabi}) as a function of the numerically simulated $\pi$-time of the transition, which is a function of the power in the $R_\perp$ beam. We see that the predicted Rabi frequencies converge at low beam powers, validating Eqs.~\eqref{4photon_rabi} and~\eqref{6photon_rabi} and suggesting that adiabatic elimination of the intermediate $\mathrm{D_{5/2}}$ levels is valid in this regime (see Fig~\ref{fig:adiabatic_approx}).


\section{Modeling of Spin Decoherence}
\label{appendix:decoherence}

To model spin decoherence we consider the effects of both slow and fast ambient magnetic field fluctuations. We model fast frequency noise as a simple exponential damping of the Rabi oscillation amplitude in time parameterized by damping rate $\gamma$. We model decoherence induced by low frequency noise as a shot-to-shot static qubit frequency offset $\delta$ drawn from a normal distribution with a width in frequency space fixed from a fit to experimental Ramsey time-series data $\sigma^{(i)}_f=1/\sigma^{(i)}_t$.

\begin{figure}[h]
\centering
    \hspace*{-0.0in}
    \includegraphics[width = 0.7\columnwidth]{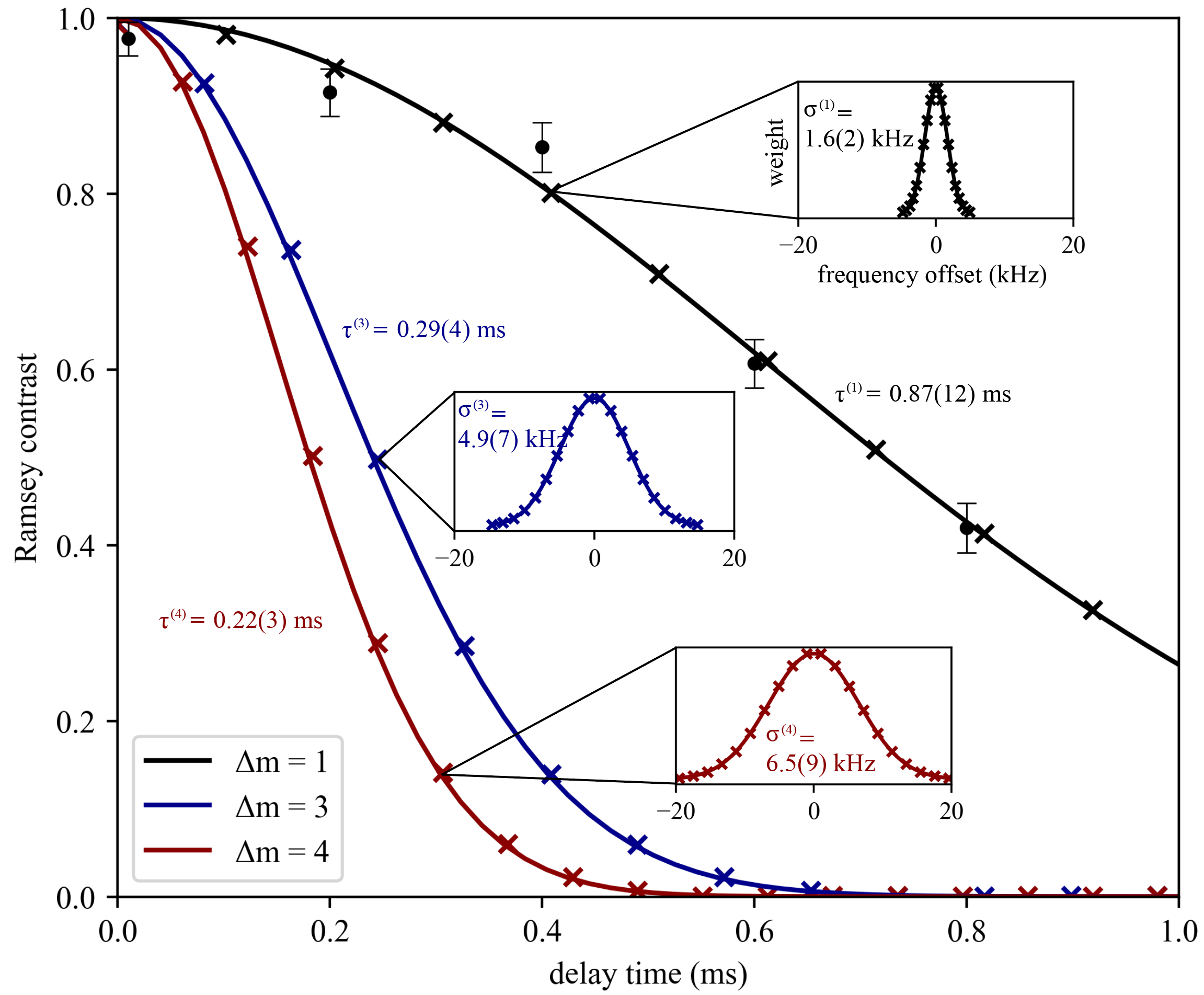}
    \caption{Ramsey contrast as a function of delay time between two $\pi/2$ pulses on the $\mathrm{m_J} =+5/2\leftrightarrow+3/2$ transition (black circles) with a Gaussian function fit to data (black line). Blue and red lines are projected contrast decay on $+5/2\leftrightarrow-1/2$ and $+5/2\leftrightarrow-3/2$ transitions, respectively. $\times$ marks are results of numerically simulated Ramsey sequences under a normally distributed shot-to-shot static qubit frequency offset with distributions shown in the insets.}
\label{fig:spincoherence}
\end{figure}

 We probed the coherence between states $\mathrm{m_J}=+5/2$ and $\mathrm{m_J}=+3/2$ by measuring the decay in Ramsey contrast (without a spin echo) as a function of delay time (see Fig.~\ref{fig:spincoherence}) using rf qubit manipulations, and expect the loss in Ramsey contrast to be dominated by the low frequency noise. The Ramsey contrast (black circles) fits well to a Gaussian decay function (solid black) with a fit standard deviation of $\sigma^{(1)}_t=0.61(8)$\,ms corresponding to a $1/e$ coherence time $\tau^{(1)}=\sqrt{2}\sigma^{(1)}_t=0.87(12)$\,ms. Numerical simulations (black $\times$ markers) assuming the corresponding $\sigma^{(1)}_f=1.6(2)$\,kHz agree well with both the experimental data and analytic Gaussian decay function. As the presented four- and six-photon transitions cover $\Delta \mathrm{m_J}=3$ and 4, respectively, we assume $\sigma^{(3)}_t=\sigma^{(1)}_t/3$ (solid blue) and $\sigma^{(4)}_t=\sigma^{(1)}_t/4$ (solid red).

Because off-resonant Rabi oscillations have a closed analytic expression, we can write the functional form of Rabi oscillations in our model as a weighted average of detuned Rabi oscillations where $\Omega$ is the Rabi frequency and $\gamma$ is the heuristic damping constant to account for loss in contrast due to faster noise components not seen in the slow noise dominated Ramsey data:
\begin{align}
    \label{decoheredflopping}
    P(t,\Omega,\sigma_f,\gamma)=\frac{1}{\sigma\sqrt{2\pi}}\sum_\delta\frac{\Omega}{\sqrt{\Omega^2+\delta^2}}\sin^2{\frac{t\sqrt{\Omega^2+\delta^2}}{2}}e^{-\frac{\sigma^2}{2\delta^2}-\gamma t}.
\end{align}

\begin{figure}[h]
\centering
    \hspace*{-0.0in}
    \includegraphics[width = \columnwidth]{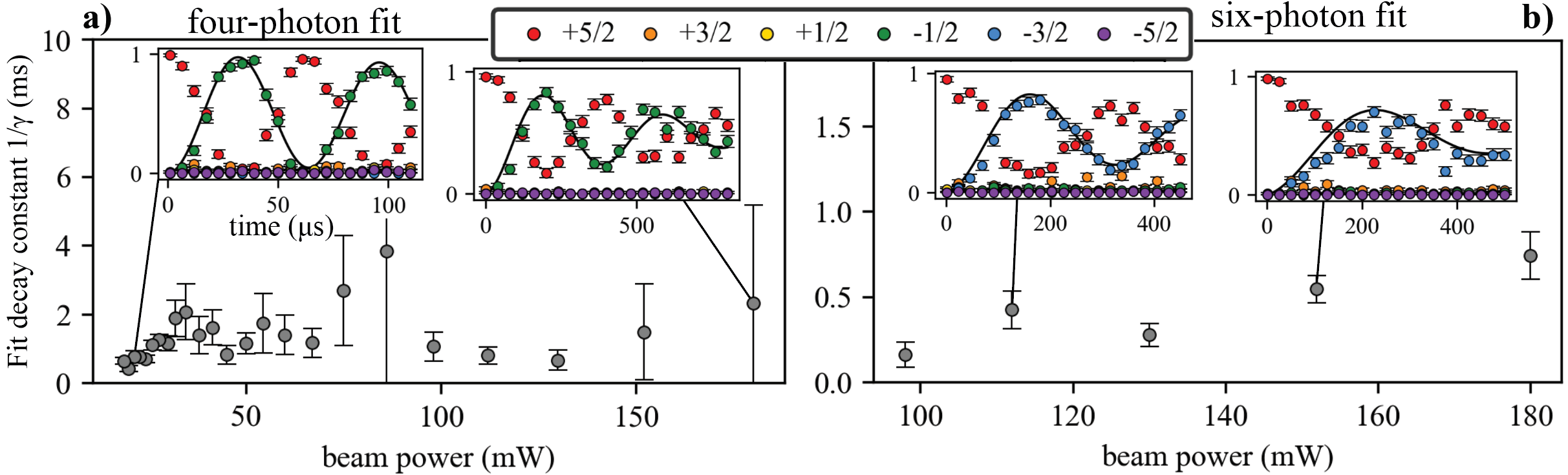}
    \caption{Returned fit parameter $\gamma$ for each set of resonant Rabi-oscillations for four (\textbf{a}) and six (\textbf{b}) photon driven time dynamics.}
\label{fig:floppingfitting}
\end{figure}

For each data set, we fit $\Omega$ and $\gamma$ in Equation~\eqref{decoheredflopping} while holding $\sigma_f$ fixed. Shown in Fig.~\ref{fig:floppingfitting} are the fits to the damping term $\gamma$ for each data set, with example Rabi oscillations in the insets. The $\Omega$ results are presented in Fig.~\ref{fig:powerscaling}b in the main text.

\section{Beam polarization and intensity calibration}
\label{appendix:calibrations}
\begin{figure}[h]
\centering
    \hspace*{-0.0in}
    \includegraphics[width = \columnwidth]{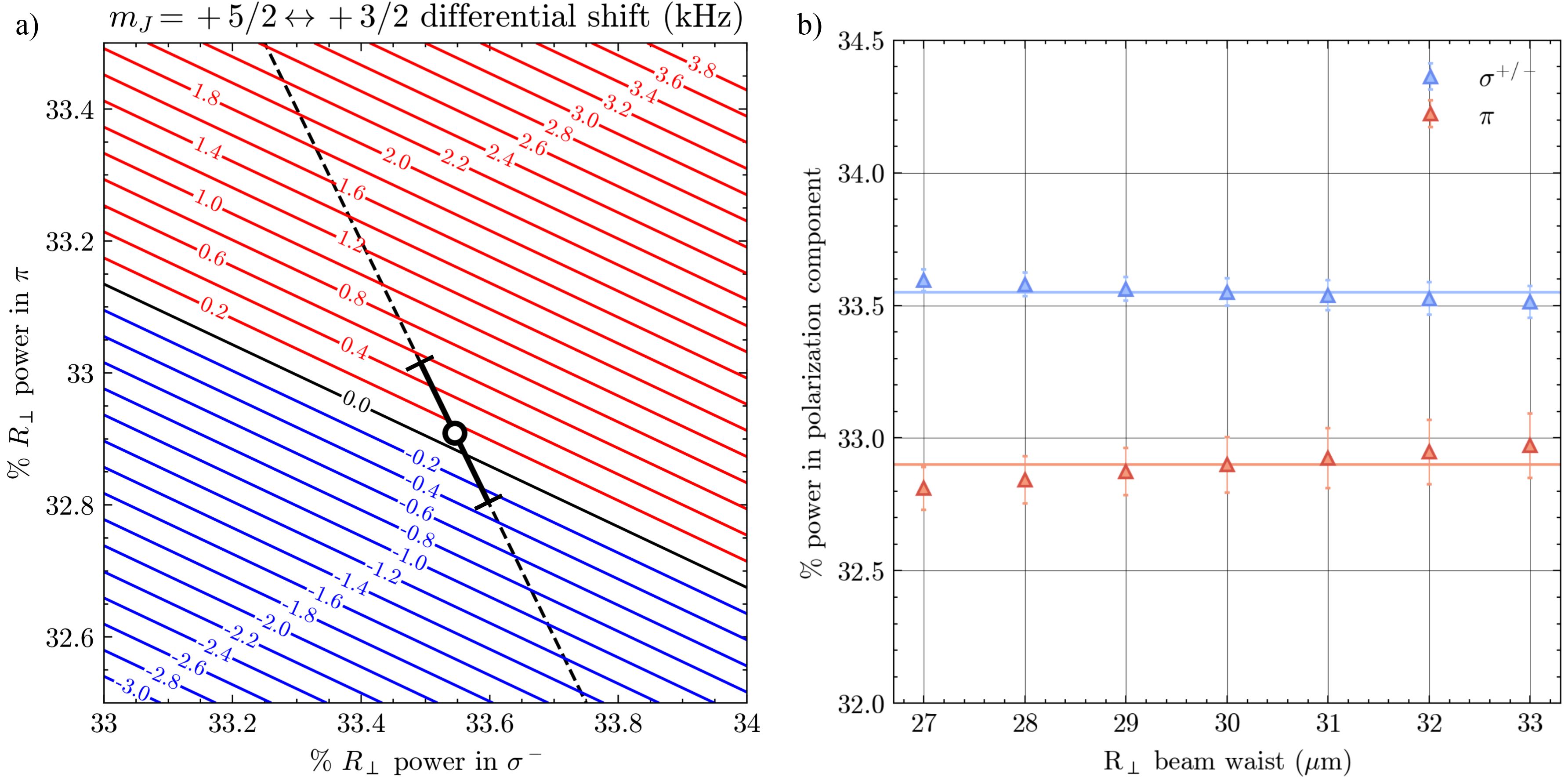}
    \caption{\textbf{a)} Differential a.c. Stark shift (kHz) up to eighth order in perturbation on the $\mathrm{m_J} = +5/2\leftrightarrow+3/2$ qubit with $R_\perp$ beam illuminating the ion at 180\,mW, assuming at 30\,\textmu m beam waist. Dashed line represents the constraint that $\sigma^{-/+}$ polarizations have the same relative intensities. \textbf{b)} Constrained $R_\perp$ beam polarization components as a function of reasonable beam waists in the experiment.}
\label{fig:rpi_constraint}
\end{figure}

The analytic expressions for the presented four- and six-photon transition Rabi frequencies (Equations~\eqref{4photon_rabi} and~\eqref{6photon_rabi}) rely on precise calibration of the intensity of each polarization component of the $R_\perp$ and $R_\parallel$ beams. To carry out this calibration, we performed a series of auxiliary experiments and fit intensities to the experimental data. At the end of this section we present a table summarizing the constrained parameters for the two beams.
\newline
\newline
\textit{$R_\perp$ polarization constraint}
\newline
\newline
Nominally, the $R_\perp$ beam has equal components of $\sigma^-/ \pi / \sigma+$ polarizations to null the lowest (second) order differential a.c. Stark shifts throughout the $\mathrm{D_{5/2}}$ manifold. We assume the beam propagates perpendicular to the quantization axis and thus has equal power in the $\sigma^-$ and $\sigma^+$ polarization components. We estimate the $\pi$ component by measuring the differential a.c. Stark shift imparted on the $\mathrm{m_J} = +5/2 \leftrightarrow+3/2$ transition by 180\,mW of the $R_\perp$ beam. We measure the unperturbed Zeeman and $R_\perp$ shifted splittings via rf Rabi spectroscopy to be 2.6338(3) and 2.6339(3) MHz, respectively, corresponding to a differential shift of 0.1(4)\,kHz. To obtain an estimate for the bound on the beams polarization makeup we calculate the differential a.c. Stark shift on the transition as a function of relative beam power in $\pi$ and $\sigma^+$ polarization up to 8th order in perturbation theory~\cite{shiftsInPrep} assuming a 30\,\textmu m beam waist (see Figure~\ref{fig:rpi_constraint}), and compare to the value we measure experimentally.

\begin{figure}[h]
\centering
    \hspace*{-0.0in}
    \includegraphics[width = .7\columnwidth]{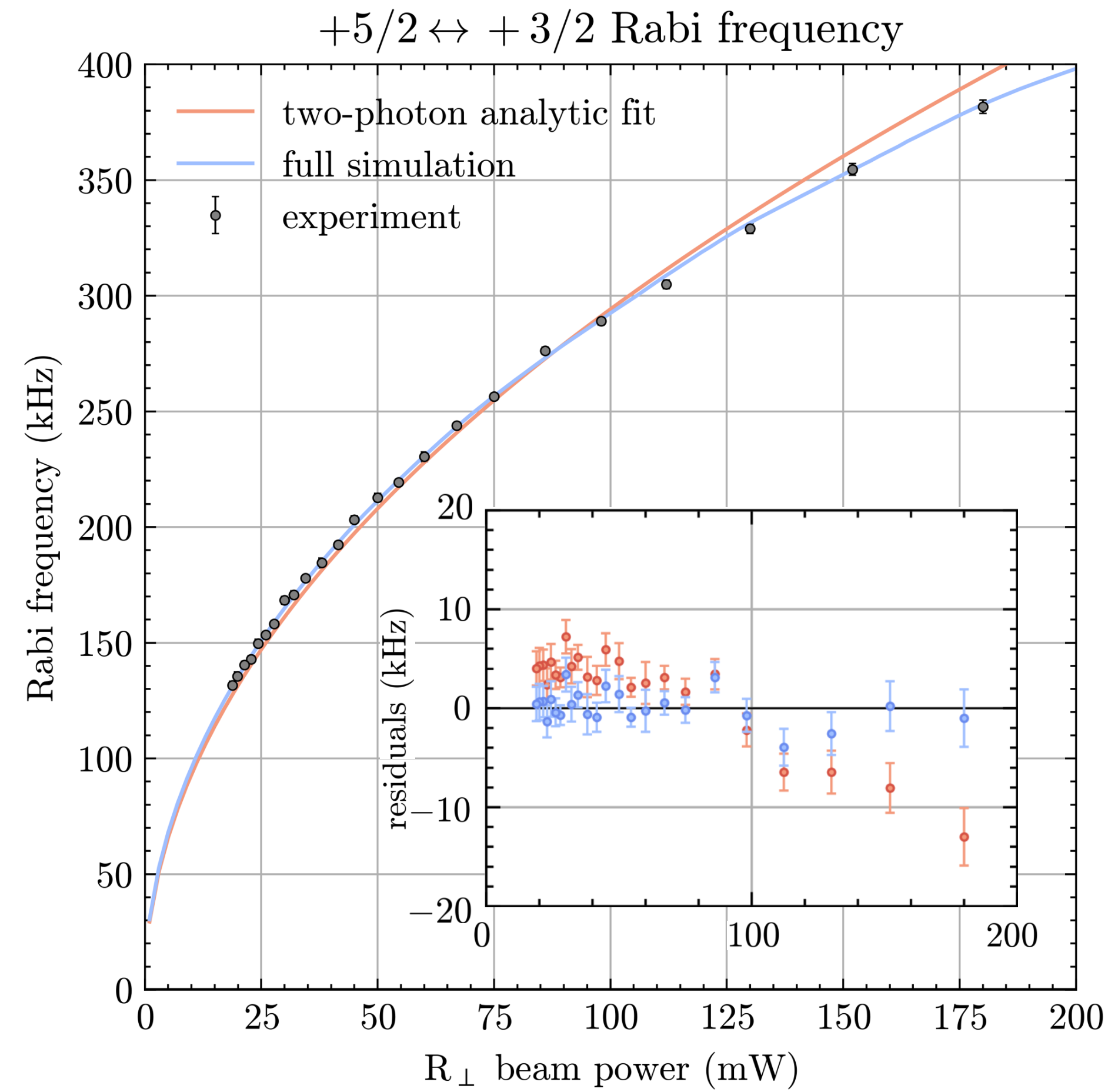}
    \caption{Rabi frequency of the $\mathrm{m_J} = +5/2 \leftrightarrow+3/2$ transition as a function of power in the $R_\perp$ beam, with the power in $R_\parallel$ held fixed at 195\,mW. Red curve is a fit to the analytic two-photon Rabi frequency expression. Blue curve is a fit to a full numerical simulation of the system natively including higher order resonant four-photon corrections to the Rabi frequency, with residuals shown in inset. }
\label{fig:two_photon_scaling}
\end{figure}

\begin{figure}[h]
\centering
    \hspace*{-0.0in}
    \includegraphics[width = \columnwidth]{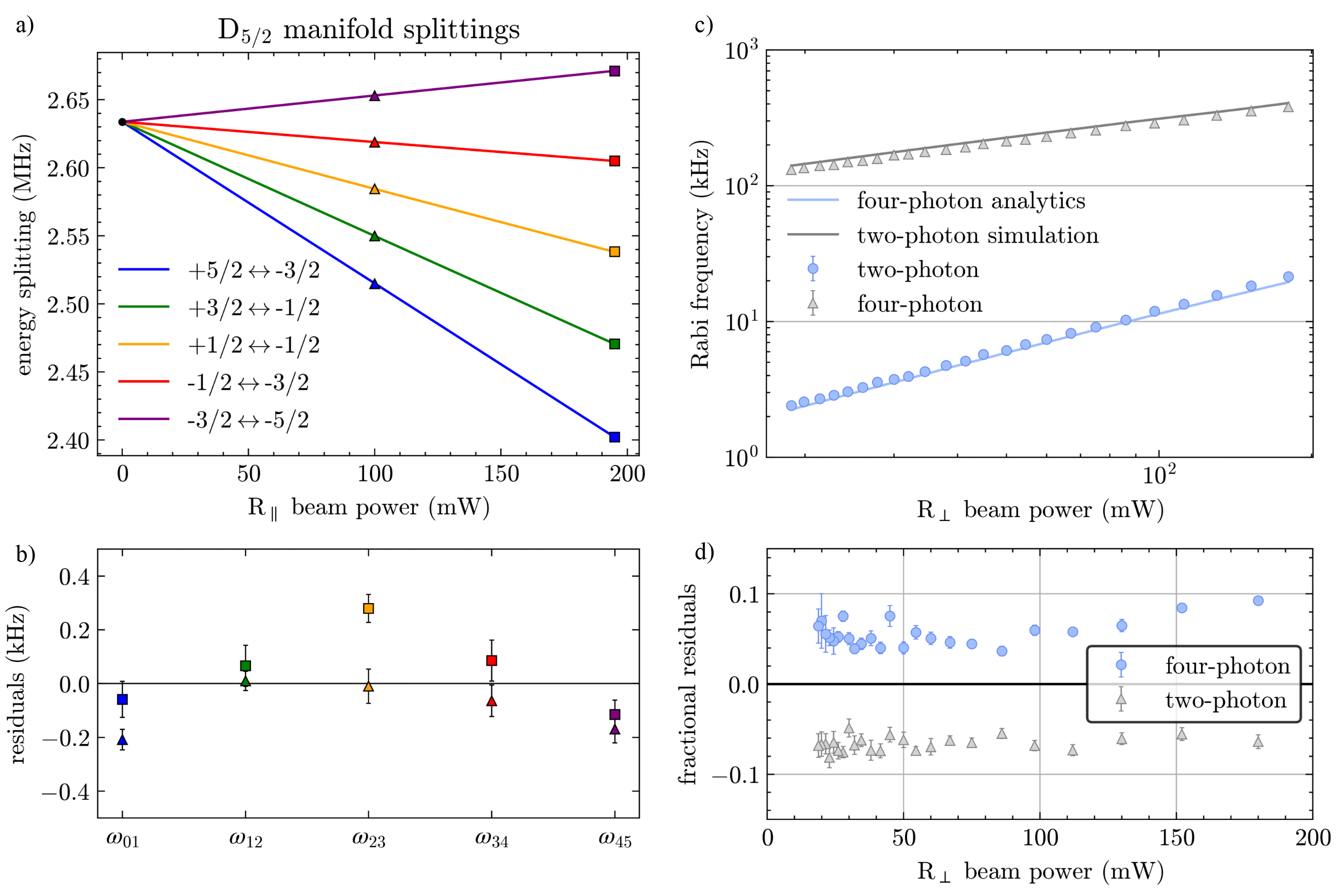}
    \caption{\textbf{a)} Experimentally measured nearest-neighbor state splittings within the $\mathrm{D_{5/2}}$ manifold at various $R_\parallel$ beam powers. Solid curves are fits to analytic expressions for the differential shift up to and including eighth order corrections. \textbf{b)} Residuals to fit shown in a).\textbf{c)} Numerically simulated Rabi frequency of the $\mathrm{m_J} = +5/2 \leftrightarrow+3/2$ transition (grey) as a function of power in the $R_\perp$ beam, with the power in $R_\parallel$ held fixed at 195\,mW. Blue curve is analytic prediction for four-photon Rabi frequency, including counter-rotating terms, coupling to $F$ states, and six-photon correction terms. \textbf{d)} Residuals to fit shown in c).}
\label{fig:rsig_constraint}
\end{figure}

We constrain the polarization breakdown of the $R_\perp$ beam to be 33.55(5)\% $\sigma^{+/-}$ and 32.9(1)\% $\pi$ (see Figure~\ref{fig:rpi_constraint}a) based on the experimentally measured differential shift. We also find polarization breakdowns that agree within error over a large range of beam waist sizes relative to the experiment Figure ~\ref{fig:rpi_constraint}(b).
\newline
\newline
\textit{$R_\parallel$ polarization and beam waist constraints}
\newline
\newline
Nominally, the $R_{\parallel}$ beam is purely $\sigma^-$ polarized. To estimate the impurity in the polarization of the beam we measured the frequency splittings between nearest-neighbor states in the $\mathrm{D_{5/2}}$ manifold under increasing $R_\parallel$ beam powers. To determine the splitting frequency between states $\ket{i}$ and $\ket{i+1}$ we prepare the ion in the $\ket{i}$ state and perform Rabi spectroscopy by driving with RF and measuring population in the $\ket{i+1}$ state while scanning the RF frequency. Shown in Fig.~\ref{fig:rsig_constraint}a are the measured nearest neighbor splittings in $\mathrm{D_{5/2}}$ at increasing $R_\parallel$ beam power. We can then write out analytic expressions for the splittings considering ac Stark shifts induced by the $R_\parallel$ beam coupling to $F_{7/2}$ and $\mathrm{P_{3/2}}$ states up to and including eighth order in perturbation ~\cite{shiftsInPrep} and counter-rotating terms, parameterized by the intensity in each polarization component. The three free parameters of the $R_\parallel$ beam are under constrained by the splittings alone, and we additionally measure the Raman Rabi frequency of the $\mathrm{m_J}=+5/2\leftrightarrow+3/2$ transition at the two-photon resonance condition and with increasing $R_\perp$ beam powers while the $R_\parallel$ beam power held constant at 195\,mW. Initially, we sought to fit the analytic expression of the well-known two-photon Rabi frequency to the data but found that the experimental measurements of the Rabi frequency did not follow the expected $\sqrt{P_\perp}$ scaling within error. We attribute this discrepancy to higher order resonant four-photon processes driving the transition simultaneously, with expected four-photon pathways contributing on the $\sim 10\%$ level. We numerically simulate the transition (blue curve in Fig~\ref{fig:two_photon_scaling}) and find matching scaling of the simulated and experimentally obtained Rabi frequency with $R_\perp$ power, suggesting that the higher order processes are captured by the simulation, and are not well characterized by the two-photon power scaling law (red curve). 

Additionally, we did not control the relative phase difference of the two Raman beams in the experiment from shot-to-shot, leaving an extra parameter for us to determine. While this does not have an effect on the usual two-photon stimulated Raman transition Rabi frequencies, different higher-order pathways can make use of different numbers of photons from each beam, meaning that the same physical phase differential between beams can lead to different phases of higher order pathways that then interfere differently depending on their phase difference. This can change the total Rabi frequency when the number of photons exceeds $\Delta\mathrm{m}$.


We performed a joint least-squares fit of the data shown in Fig~\ref{fig:rsig_constraint} to analytic predictions of the nearest neighbor energy splittings four-photon Rabi frequencies and numerical simulations of the two-photon Rabi frequency. We floated the beam waists ($w_0$) of the $R_\perp$ and $R_\parallel$ beams, the polarization vector for the $R_\parallel$ beam with $\mathbf{E} = |E|(\mathbf{e_{\sigma^-}+e_{\pi}+e_{\sigma^+}})$, and the relative phase difference of the Raman beams $\phi$. Together with the constraints on the $R_\pi$ beam polarization (Fig~\ref{fig:rpi_constraint}, this parameterization fully describes the  $R_\perp$ and $R_\parallel$ beams. We present the fit parameters in Table~\ref{tab:fit_params}.

\renewcommand{\arraystretch}{1.3}
\begin{table}
\caption{\label{tab:fit_params}Summary of fit beam waist and polarization components of the Raman beams used for theoretical four- and six-photon analytic Rabi frequencies (see Fig.~\ref{fig:powerscaling}b.)}
\setlength{\tabcolsep}{8pt}
\begin{tabular}{>{\bfseries}l | c c c c }

Parameter & $w_0$ (\textmu m) & $e_{\sigma^-}^2$ & $e_{\pi}^2$ & $e_{\sigma^+}^2$ \\[5pt]
\hline
Fit value $R_\parallel$ & 31.52(3) & 92.29(7)\% & 0.16(7)\% & 7.55(2)\%\\
Fit value $R_\perp$ & 30.0(2) & 33.55(5)\% & 32.9(1)\% & 33.55(5)\% \\
\end{tabular}
\end{table}

\clearpage

\section{Corrections to analytic Rabi frequencies}
\label{appendix:analytic_corrections}

\subsection{$\mathrm{F}$ state coupling}
The $976$\,nm beams we use are $-44$\,THz red-detuned from the $854$\,nm $\mathrm{D_{5/2}}\leftrightarrow \mathrm{P_{3/2}}$ transition. The beams are also $-1,322$\,THz red-detuned from the $184$\,nm $\mathrm{D_{5/2}}\leftrightarrow \mathrm{F_{7/2}}$ and $\mathrm{D_{5/2}}\leftrightarrow \mathrm{F_{5/2}}$ transitions~\cite{UDportal,NIST}. In this section, we discuss small corrections to Equations~\eqref{4photon_rabi} and~\eqref{6photon_rabi} considering couplings to the $\mathrm{F}$ manifolds. The matrix elements coupling manifolds $3\mathrm{D_{5/2}}\leftrightarrow4\mathrm{P_{3/2}}$, $3\mathrm{D_{5/2}}\leftrightarrow4\mathrm{F_{5/2}}$, and $3\mathrm{D_{5/2}}\leftrightarrow4\mathrm{F_{7/2}}$ are (in atomic units) 3.283(6), 0.5164(56), and 2.309(25) respectively, so we ignore couplings to $\mathrm{F_{5/2}}$ due to the relatively small matrix element.

\begin{figure}[h]
\centering
    \hspace*{-0.0in}
    \includegraphics[width = 0.6\columnwidth]{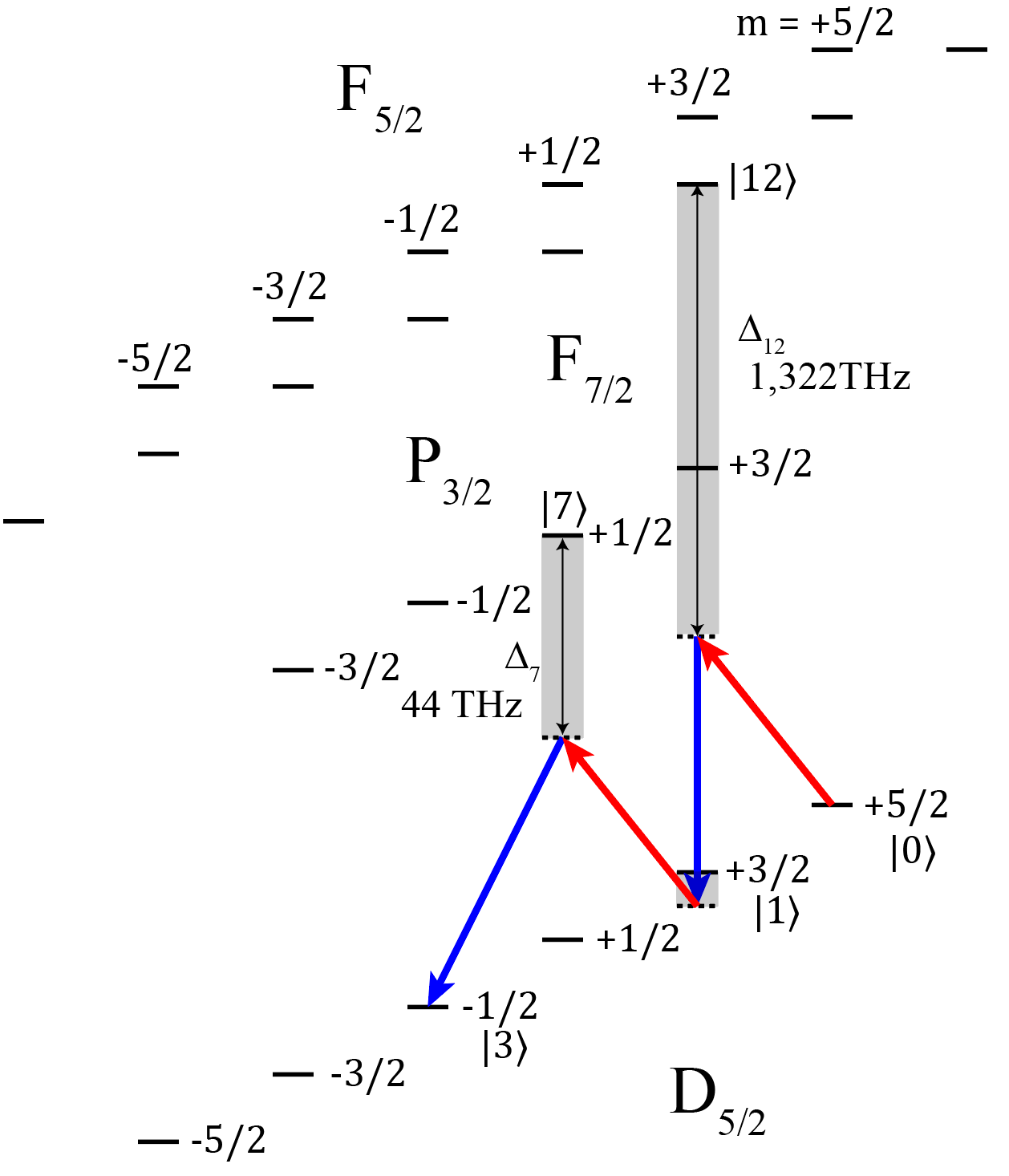}
    \caption{Example resonant four-photon pathway from $\mathrm{m_J}=+5/2$ to $\mathrm{m_J}=-1/2$ in $\mathrm{D_{5/2}}$ with couplings to excited states $\mathrm{F_{7/2}},\mathrm{m_J} = +3/2$ and $\mathrm{P_{3/2}},\mathrm{m_J}=+1/2$. The corresponding contribution to the four-photon Rabi frequency is given in Equation~\eqref{f_term}. Other resonant pathways are not illustrated.}
\label{fig:F_state_coupling}
\end{figure}
There are many pathways involving these manifolds that facilitate the $+5/2 \leftrightarrow-1/2$ four-photon transition. We neglect terms where the $\mathrm{F}$ states are coupled to more than once in a given pathway, as the corresponding term for the Rabi frequency contains the larger $\Delta_F$ squared (or cubed) in the denominator. Pathways coupling to both the $\mathrm{F}$ and $\mathrm{P}$ states, however, contribute percent-level effects. In Fig.~\ref{fig:F_state_coupling} we show an example pathway for the four-photon transition we present in the main text including coupling to the $\mathrm{F_{7/2}}$ manifold. This pathway adds the term
\begin{align}
       \frac{\Omega^*_{0 12}\Omega_{1 12}\Omega^*_{17}\Omega_{37}}{8\Delta_{12}\Delta_7(\omega_r-\omega_{01})}
       \label{f_term}
\end{align}
to the four-photon Rabi frequency (Equation~\eqref{4photon_rabi}). Corresponding terms for all relevant pathways for both the four- and six-photon transitions  are included in the analytic results presented in Fig.~\ref{fig:powerscaling}.

\subsection{Counter-rotating terms}
We presented the derivation of the higher order Rabi frequencies beginning with Equation~\eqref{couping}, which describes the interaction of the laser-field with the atom. Equation~\eqref{couping} is however an approximation, valid in the near-detuning regime, wherein the single beam energy $\hbar\omega$ is close to the energy splitting between the two states $E$. In this sections we derive corrections to the higher order Rabi frequencies that appear outside of this approximation.

\begin{figure}[h]
\centering
    \hspace*{-0.0in}
    \includegraphics[width = 0.5\columnwidth]{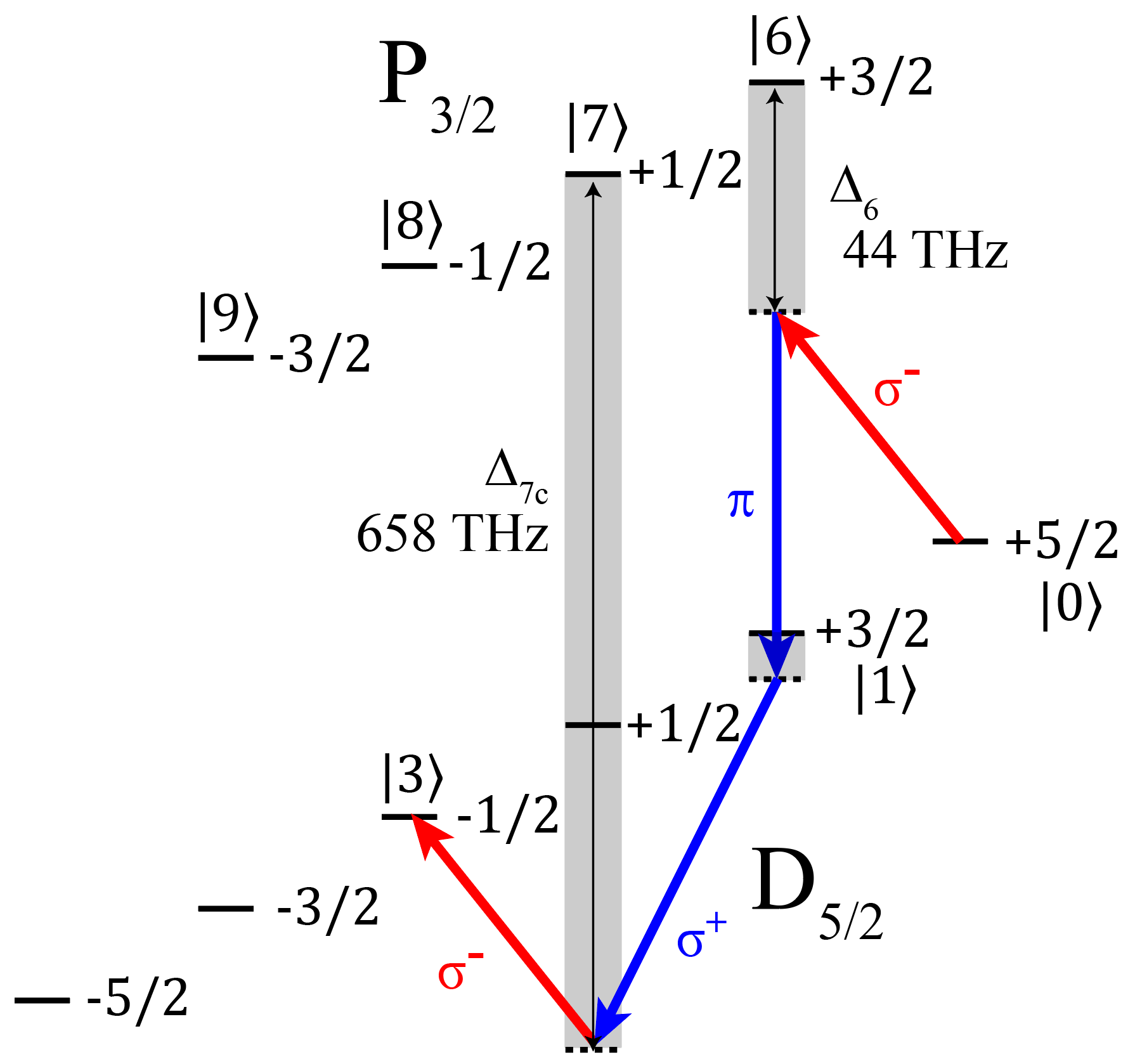}
    \caption{Example resonant four-photon pathway from $\mathrm{m_J}=+5/2$ to $\mathrm{m_J}=-1/2$ in $\mathrm{D_{5/2}}$. The two-photon pathway coupling $\mathrm{m_J}=+3/2\leftrightarrow-1/2$ is a counter-rotating pathway, coupling to $\mathrm{m_J}=+1/2$ in $\mathrm{P_{3/2}}$, far detuned by $658$\,THz. The corresponding contribution to the four-photon Rabi frequency is given in Equation~\eqref{c_term}. Other resonant pathways are not illustrated.}
\label{fig:counter_rotating_terms}
\end{figure}

A laser-field can be described by its electric field
\begin{align}
\mathbf{E_r}=\mathbf{\epsilon_r}|E_r|\cos\left(\mathbf{k_r\cdot r}-\omega_rt\right)
    \label{E1}
\end{align}
with amplitude $E_r$, k-vector $\mathbf{k_r}$, and polarization vector $\mathbf{\epsilon_r}$. The atom-laser interaction Hamiltonian is given by the dot product of the electric field and the dipole moment of the atom $\mathbf{\mu}$
\begin{align}
    V=-\mathbf{\mu}\cdot \mathbf{E}=e\mathbf{r\cdot \mathbf{E}}
\end{align}
with $e$ the fundamental unit charge, and $\mathbf{r}$ the position operator. We can now calculate the $ij$'th component of $V$:
\begin{align}
    V_{ij}=\bra{i}V\ket{j}=e\bra{i}\mathbf{r}\cdot \mathbf{E}\ket{j}
\end{align}
Plugging in Equation~\eqref{E1} and expanding the cosine:
\begin{align}
    V_{ij}&=\frac{e|E_r|}{2}\bra{i}\mathbf{r}\cdot\mathbf{\epsilon_r}\ket{j}\left(e^{i(\mathbf{k_r}\cdot \mathbf{r} - \omega_rt)} + e^{-i(\mathbf{k_r}\cdot \mathbf{r} - \omega_rt)}\right)
\end{align}
Which, if we define the single beam Rabi frequency:
\begin{align}
    \Omega_{ij}\equiv  \bra{i}\mathbf{r}\cdot\mathbf{\epsilon_c}\ket{j}\frac{e|E_c|}{\hbar}e^{i\mathbf{k_c}\cdot \mathbf{r}}
\end{align}
reduces to
\begin{align}
    V_{ij}&=\Omega_{ij}e^{-i\omega_rt} + \Omega^*_{ij}e^{i\omega_rt}
    \label{full_coupling}
\end{align}
Equation~\eqref{couping} only contains the term with the positive frequency component. This is because after transforming to the interaction representation with respect to the bare atom energies, these terms transform to (for $i = 0$ and $j = 6$):

\begin{align}
    V_{ij}&=\Omega_{ij}e^{-i(\omega_r + E_{06}/\hbar)t} + \Omega^*_{ij}e^{i(\omega_r-E_{06}/\hbar)t}
\end{align}
Equation~\eqref{couping} is valid in the near-detuned regime, assuming $\omega_r + E_{06}/\hbar\gg\omega_r - E_{06}/\hbar$, and the term with $\omega_r + E_{06}$ in the argument of the complex exponential is ignored in a RWA. Keeping both terms in Equation~\eqref{full_coupling}, we can derive corrections to the higher order Rabi frequencies that involve these counter-rotating terms. Shown in Figure~\ref{fig:counter_rotating_terms} is an example pathway involving these corrections. For the $976$\,nm beams used in this work the $\mathrm{P}$ state detuning via a counter rotating path is $\sim658$\,THz, which compared to the $44$\,THz $\mathrm{P}$ state detuning of the nominal terms contribute percent level corrections to the Rabi frequency.

The pathway shown in Figure~\ref{fig:counter_rotating_terms} corresponds to a correction term:
\begin{align}
       \frac{\Omega^*_{0 6r}\Omega_{17b}\Omega^*_{17bc}\Omega_{37rc}}{8\Delta_{6}\Delta_{7c}(\omega_r-\omega_{01})}
       \label{c_term}
\end{align}
where $\Delta_{7c}$ is the $\mathrm{P}$ state detuning from the counter-rotating two-photon coupling term. We include all counter-rotating pathways to the analytics presented in Figure~\ref{fig:powerscaling}b, and include the counter-rotating two-photon couplings in the numerical simulations.

\subsection{Six-photon corrections}
\begin{figure}[h]
\centering
    \hspace*{-0.0in}
    \includegraphics[width = 0.99\columnwidth]{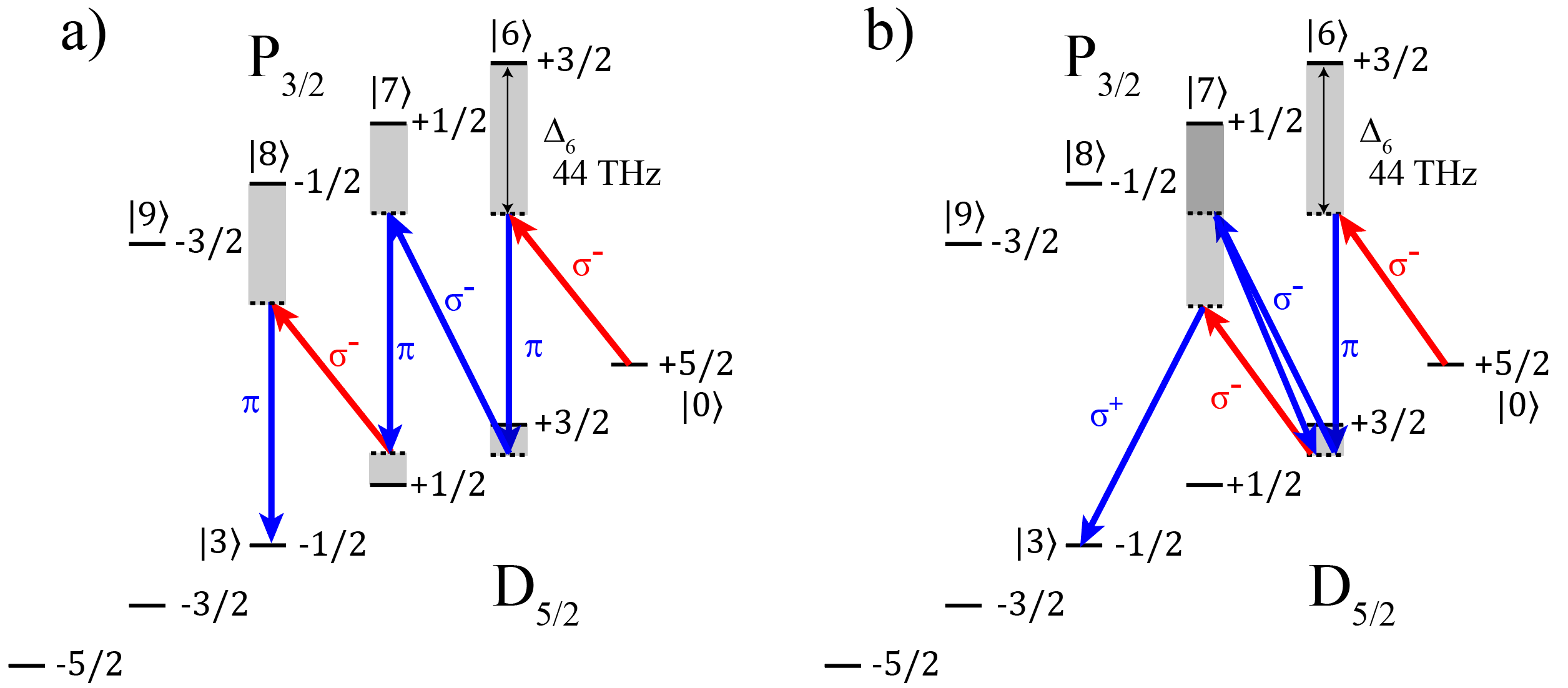}
    \caption{\textbf{a)} Example resonant six-photon pathway correction to the four-photon Rabi frequency. \textbf{b)} Resonant six-photon pathway correction. This pathway includes a photon pair ``doubling back," and terms of this form are not included in the six-photon corrections, as guided by numerical simulations.}
\label{fig:six_photon_terms}
\end{figure}

The resonance condition we set (Equation~\eqref{resonance}) to drive the four-photon transition also enables higher order resonant six-photon pathways, involving two red photons and four blue photons, or four red photons and two blue photons. We show example six-photon correction pathways to the four-photon transition in Figure~\ref{fig:six_photon_terms}a. 

The correction term to the Rabi frequency associated with the pathway drawn in Figure~\ref{fig:six_photon_terms}a is:
\begin{align}
       \frac{\Omega^*_{0 6r}\Omega_{16b}\Omega^*_{17b}\Omega_{27b}\Omega_{28r}\Omega_{38b}}{32\Delta_{6}\Delta_{7}\Delta_{8}(\omega_r-\omega_{01})(\omega_r-\omega_{01}-\omega_{12})}
       \label{six_term}
\end{align}
We found that when including six-photon pathways of the form shown in Figure~\ref{fig:six_photon_terms}b the analytic and numerically simulated Rabi frequencies do not converge. As such, we did not include six-photon corrections that include photon-pairs ``doubling back" at some point during the pathway. Included in the analytic predictions are the second order a.c. Stark shifts to the intermediate states, and we suspect that including terms of the form of Figure~\ref{fig:six_photon_terms}b may be double counting those pathways, which are already taken into account by including the second order Stark shifts.

\begin{figure}[h]
\centering
    \hspace*{-0.0in}
    \includegraphics[width = 1\columnwidth]{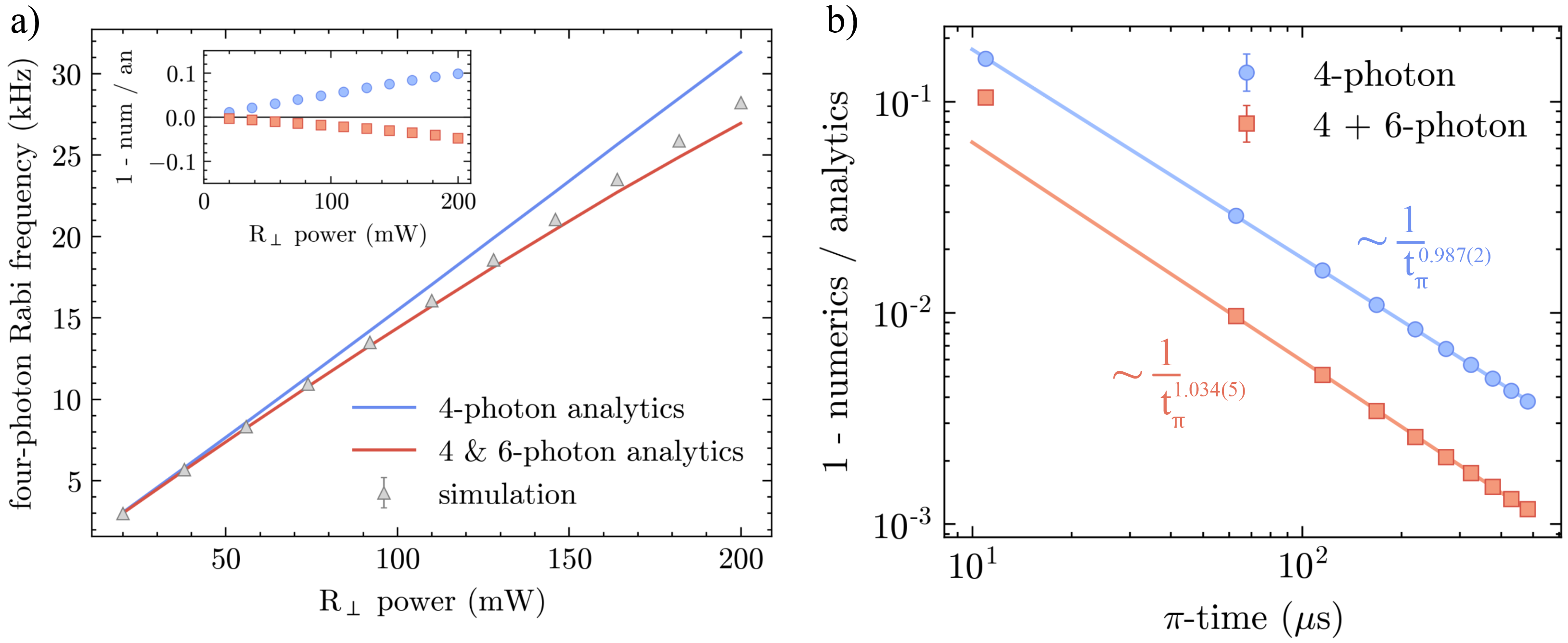}
    \caption{\textbf{a)} Numerically simulated four-photon Rabi frequency (triangles) as a function of power in the R$_\perp$ beam. Blue and red curves are analytic predictions with (red) and without (blue) six-photon correction terms. Inset is fractional residual between simulation and analytics with (red) and without (blue) six-photon corrections. \textbf{b)} Ratio of numeric to analytic predictions of the Rabi frequency with (red) and without (blue) six-photon corrections. Both analytic predictions converge with simulation, with faster convergence when including six-photon corrections.}
\label{fig:six_photon_corrections}
\end{figure}

The four-photon analytics diverge from numerical simulation as the power increases (see Figure~\ref{fig:six_photon_corrections}). When adding in the six-photon correction analytics, we find that analytics and simulation agree better over a longer range of beam powers. We expect that further disagreement between the analytics (including six-photon corrections) and simulation is due to both higher order eight-photon processes that we do not consider, and the breakdown of the adiabatic elimination approximation. 


\section{Intermediate state population}
\label{appendix:intermediatestates}
\begin{figure}[h]
\centering
    \hspace*{-0.0in}
    \includegraphics[width =0.8\columnwidth]{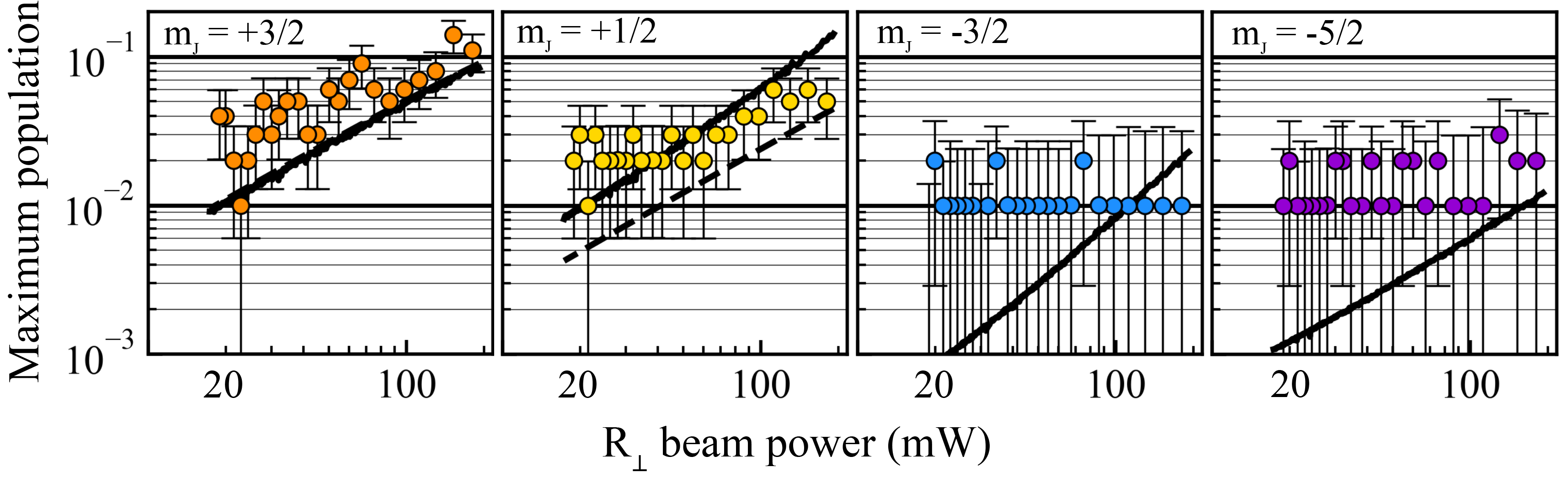}
    \caption{Maximum intermediate state population during resonant $m=+5/2 \leftrightarrow -1/2$ four-photon time dynamics at different $R_\perp$ beam powers. Solid curves are predictions from numerical simulation while dashed curves correspond to analytic ansatzes (Equations~\eqref{p32intermediatepop} and~\ref{p12intermediatepop}).}
\label{fig:intermediatestatepopulation}
\end{figure}

Numerical simulations of the four- and six-photon dynamics also provide insight into the time evolution of the intermediate state amplitudes, which are assumed to be zero in the analytic treatment. As can be seen in Fig.~\ref{fig:convergence}a, the population of the intermediate states oscillate much faster than the resonant transition Rabi frequency. We choose to place the maximum population of intermediate states during resonant dynamics as a conservative lower bound on the transfer infidelity. Shown in Fig.~\ref{fig:intermediatestatepopulation} are the maximum intermediate state populations measured in the four-photon $+5/2 \leftrightarrow -1/2$ resonant time dynamics at different beam powers. We also plot the numerically predicted bounds on the intermediate state populations resultant from the simulations (solid lines). 

The $\mathrm{m_J}=-3/2$ and $-5/2$ population bounds do not increase with beam power and remain around the measured qu$d$it SPAM error. On the other hand, the bounds for the $+3/2$ and $+1/2$ states do increase with beam power in agreement with our numerical simulations. Additionally, we posit analytic expressions for the intermediate state populations as a generalization of previous work setting a bound on intermediate state population for a three-level cascade system~\cite{intermediatestates}. 

\begin{figure}[h]
\centering
    \hspace*{-0.0in}
    \includegraphics[width = 0.8\columnwidth]{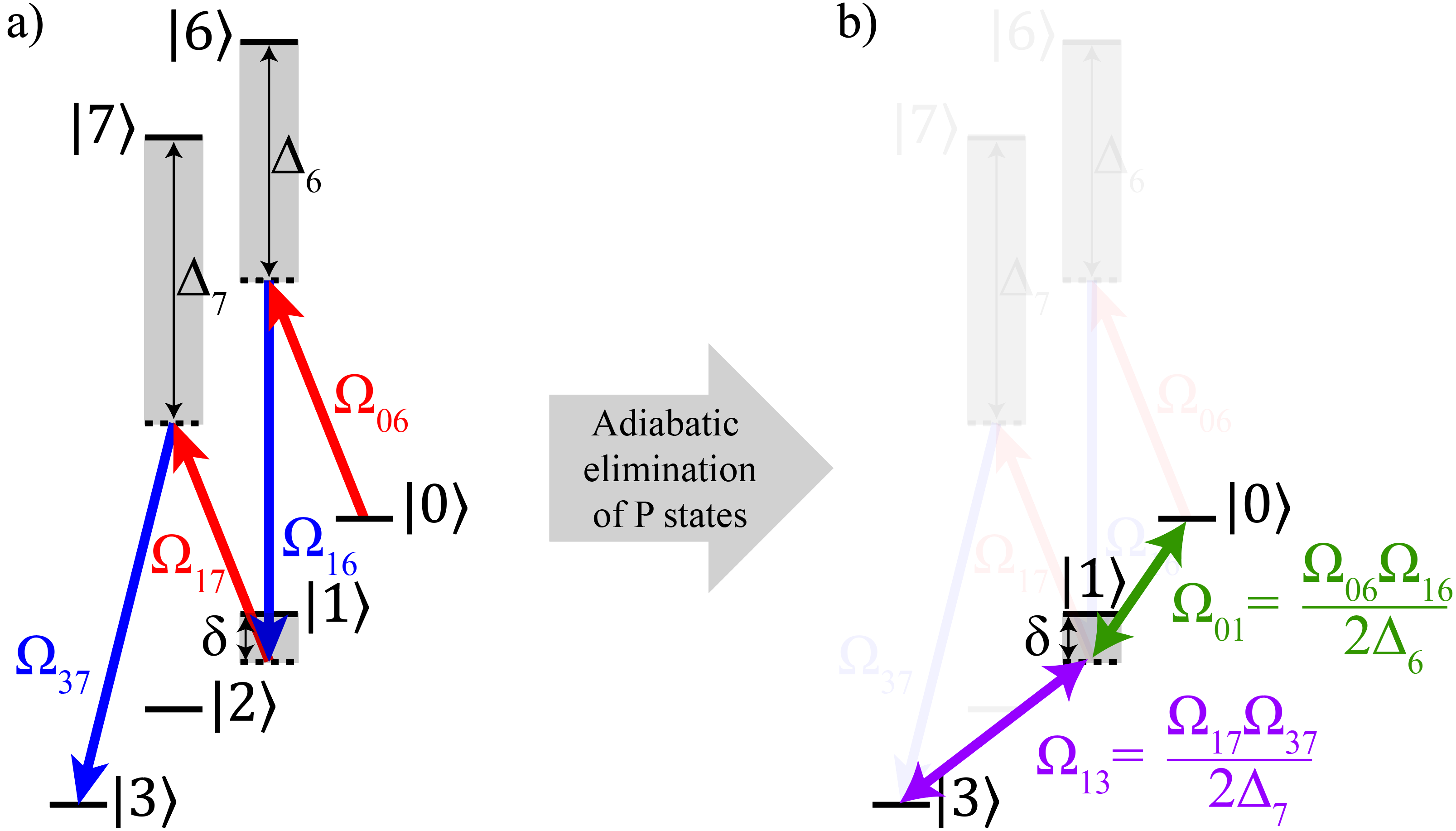}
    \caption{a). Diagram of a single resonant pathway driving a four-photon transition between angular momentum state $\mathrm{m_J}=+5/2$ and $-1/2$. b). Mapping of the 5 state system with $\mathrm{P}$ state couplings to a three level cascade system, with $\mathrm{P}$ states adiabatically eliminated.}
\label{fig:int_state_generalization}
\end{figure}

Shown in Fig.~\ref{fig:int_state_generalization}a is a single resonant pathway for the four-photon transition presented in the main text. Upon adiabatic elimination of the $\mathrm{P}$ states the five-state system is effectively simplified to a three-state cascade system for which analytic expressions for the bound on intermediate state population $|P_{1}|$ during resonant dynamics has been worked out in Ref.~\cite{intermediatestates}:
\begin{align}
    |P_{1}| = \frac{|\Omega^2_{01}|}{|\Omega^2_{01}| + |\Omega^2_{13}| + \delta^2}.
    \label{three_level_cascade}
\end{align}

We can estimate bounds on the population of intermediate states $\mathrm{m_J}=+3/2$ and $+1/2$ during the resonant four-photon process (Fig.~\ref{fig:paths_flopping}) by replacing the single beam Rabi frequencies in Equation~\eqref{three_level_cascade} with the corresponding two-beam Raman Rabi frequencies considering couplings to the $\mathrm{P}$ states. Assuming maximum populations of the intermediate states is small, the bounds for states $\mathrm{m_J}=+3/2$ and $+1/2$ are
\begin{align}
    \label{p32intermediatepop}
    |P_{+3/2}| = \frac{\Omega^2_{06}\Omega^2_{16}}{\Omega^2_{06}\Omega^2_{16} + \Omega^2_{17}\Omega^2_{37}+4\Delta^2\left(\omega_r-\omega_{01}\right)^2}
\end{align}

\begin{align}
    \label{p12intermediatepop}
    |P_{+1/2}| = \frac{\Omega^2_{06}\Omega^2_{26}}{\Omega^2_{06}\Omega^2_{26} + \Omega^2_{28}\Omega^2_{38}+4\Delta^2\left(\omega_{01}+\omega_{12}-\omega_{r}\right)^2}.
\end{align}

We plot Equations~\eqref{p32intermediatepop} and~\eqref{p12intermediatepop} as dashed lines in Fig.~\ref{fig:intermediatestatepopulation} and find reasonable agreement between the analytic predictions and numerical simulations. 


\section{Extension to high fidelity}
\label{appendix:highfid}

In this section we discuss pathways towards higher fidelity control of the four-photon transition discussed in the main text, including considerations of the dominant error sources: population of intermediate states as a breakdown of the adiabatic elimination approximation, decoherence of the qudit spin states caused by ambient magnetic field fluctuations, and errors incurred through spontaneous Raman scattering. We do not discuss the six-photon transition in detail, but similar arguments can be made to increase its fidelity as well.
\newline
\newline
\textbf{\textit{Intermediate state population}}
\newline
\label{intstatefidelity}

\begin{figure}[h]
\centering
    \hspace*{-0.0in}
    \includegraphics[width = 0.9\columnwidth]{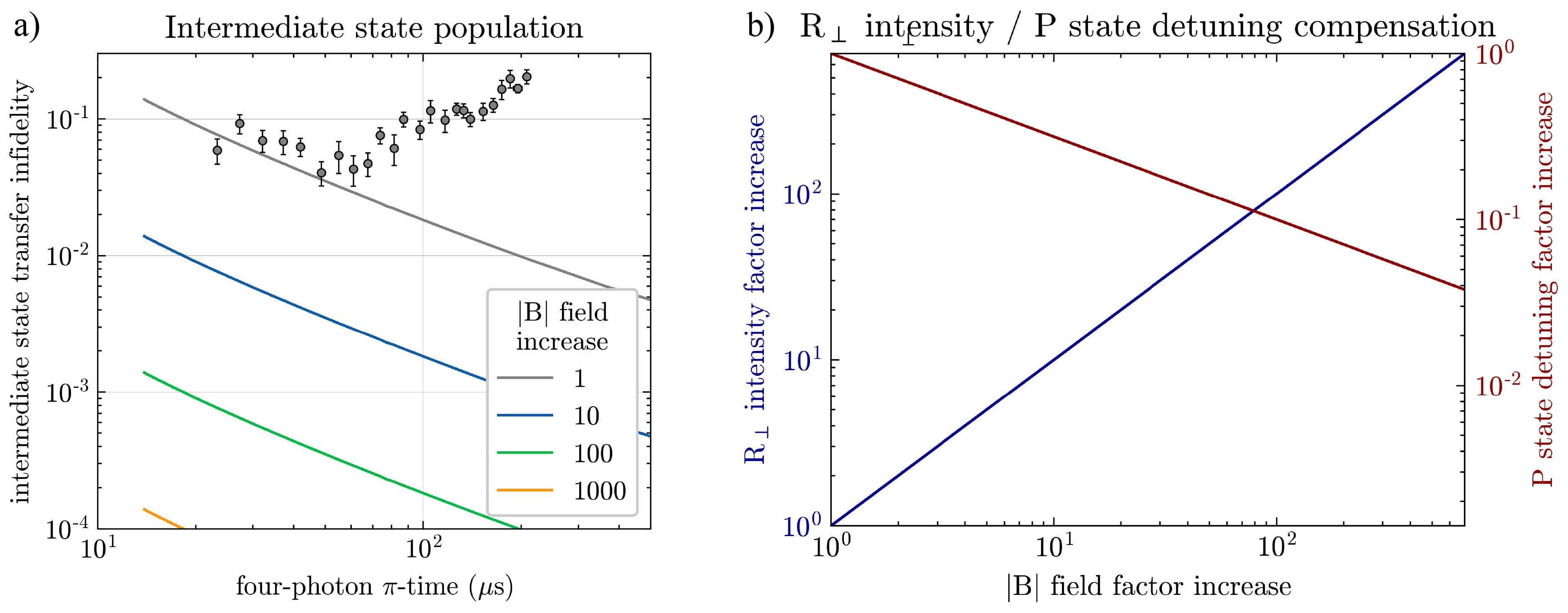}
    \caption{\textbf{a)} Four-photon $\pi$ pulse transfer infidelity manifested as population of intermediate states, determined by numerical simulations (black), and extrapolated to other Zeeman splittings according to analytic model (Equation~\eqref{p32intermediatepop}). \textbf{b)} Compensation factors of $R_\perp$ intensity or $\mathrm{P}$ state detuning to maintain constant four-photon Rabi frequency for different Zeeman splitting increases.}
\label{fig:int_state_extrapolation}
\end{figure}

One strategy to reduce intermediate state population is increasing the magnitude of the quantization magnetic field, thus increasing $\delta_1$ and $\delta_2$ (see Figure~\ref{fig:adiabatic_approx}). In principle, this strategy can reduce the associated infidelity arbitrarily. However, the reduction in Rabi frequency when increasing $\delta_1$ and $\delta_2$ must be accounted for with a corresponding increase in beam intensity or decrease in $\mathrm{P}$ state detuning, and we find that lowering the intermediate state population to $<10^{-4}$ level requires a $10^3$ factor increase in quantization field strength and beam intensity. Greater beam intensity for the same pulse length causes significantly more Raman scattering errors, making them the dominant error at 100s of Gauss.

Alternatively, we can view the population of intermediate states as a consequence of off-resonant Rabi oscillations. For a square pulse, the power spectral density (PSD) takes on a $\mathrm{sinc}^2(f)$ profile, with lobes that fall off as $1/f$ away from the carrier frequency. We consider applying an amplitude envelope to the $R_\perp$ beam to lower the PSD at the intermediate state locations in frequency space and reduce the population driven to the intermediate states~\cite{pulse_shaping}. 
We consider $\sin^2$ and $\sin^4$ amplitude ramps on the $R_\perp$ beam (while keeping the shape of the $R_\parallel$ beam pulse square) and carry out numerical simulations of the dynamics to estimate the improvement of four-photon $\pi$ pulse transfer fidelity. We maintain the same peak power as the square pulses and use ramp parameters such that the total pulse time is increased by a factor of $1.25\times$. 

\begin{figure}[h]
\centering
    \hspace*{-0.0in}
    \includegraphics[width = 0.9\columnwidth]{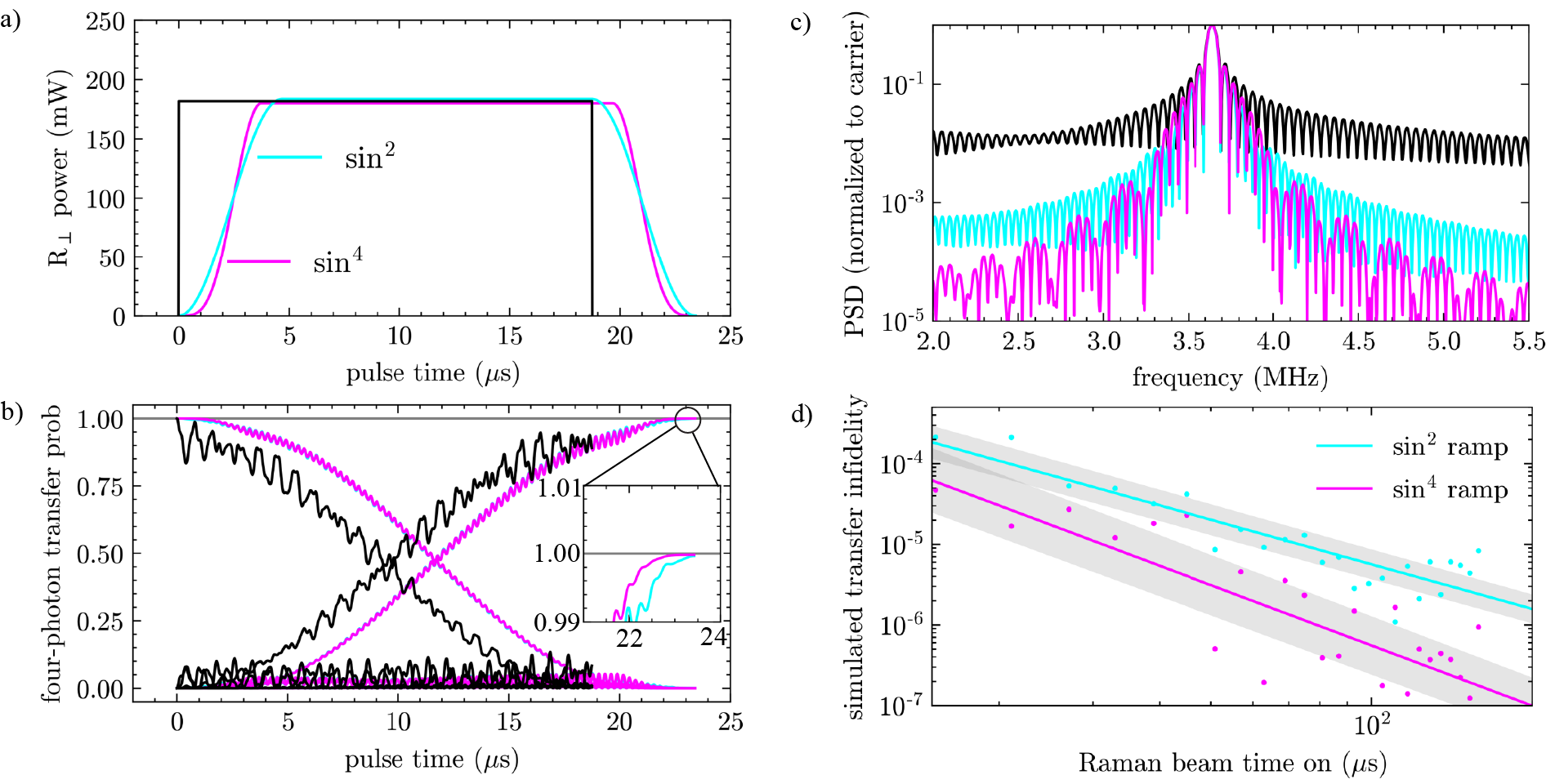}
    \caption{\textbf{a)} Amplitude envelopes for the $R_\perp$ beam. \textbf{b)} Simulated population dynamics for the four-photon transition $\pi$ pulse driven by the pulse shaping shown in a). \textbf{c)} PSD of the three pulses driving population in b) with envelopes shown in a). \textbf{d)} Simulated $\pi$ pulse transfer infidelity with the pulse shaping shown in a) at different $\pi$ times, corresponding to different $R_\perp$ peak beam powers. }
\label{fig:pulse_shape}
\end{figure}

Illustrated in Figure~\ref{fig:pulse_shape}a are the three amplitude envelopes we consider, with the peak power in $R_\perp$ and $R_\parallel$ beams set to 180 and 195\,mW, respectively, coinciding with the maximum powers used in the experiment. In Figure~\ref{fig:pulse_shape}b we show the simulated population dynamics of the full six-level system corresponding to a $\pi$ pulse with pulse shaping illustrated in Figure~\ref{fig:pulse_shape}a, with the amplitude of intermediate state population oscillations clearly diminished due to the pulse shaping, especially at the beginning and end of the pulse. The corresponding PSD of each pulse is shown in Figure~\ref{fig:pulse_shape}c, illustrating the several orders of magnitude reduction in PSD at the intermediate state locations relative to the carrier. Finally, we simulate $\pi$ pulses with pulse shaping at different powers of the $R_\perp$ beam, corresponding to different $\pi$ times. We take the final population in $\ket{3}$ after simulating a $\pi$ pulse on the system initialized in $\ket{0}$ as the transfer fidelity, and fit a straight line to the logarithms of the simulated infidelity vs $\pi$ time data shown in Figure~\ref{fig:pulse_shape}d. 

\begin{figure}[h]
\centering
    \hspace*{-0.0in}
    \includegraphics[width = 0.9\columnwidth]{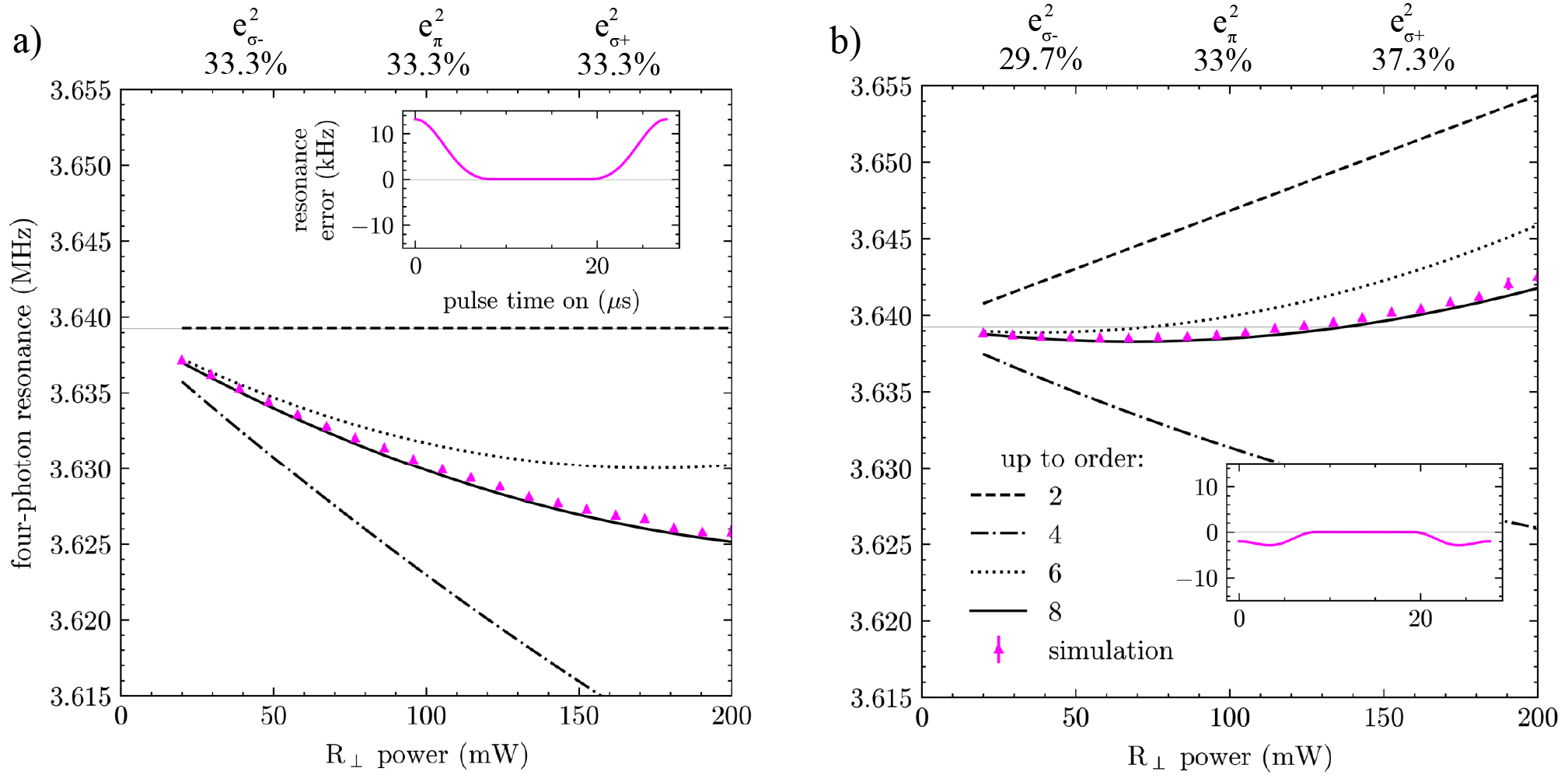}
    \caption{\textbf{a)} Simulated four-photon resonance with equal powers in each polarization component of the $R_\perp$ beam. Black curves correspond to analytic corrections from perturbation theory up to second (dashed), fourth (dot-dashed), sixth (dotted) and eighth (solid) order~\cite{shiftsInPrep}. Inset shows time-dependent shift  during the $\pi$ pulse with $sin^2$ amplitude ramps on the $R_\perp$ beam and maximum beam powers. \textbf{b)} Same as a) with $R_\perp$ polarizations chosen to reduce time-dependent shift.}
\label{fig:differential_shift}
\end{figure}

Any un-nulled differential a.c. Stark shifts between the states $\ket{0}$ and $\ket{3}$ cause the four-photon transition to shift out of resonance during amplitude ramping, leading to a small transfer infidelity. a.c. Stark shifts up to 8th order in perturbation theory contribute\,kHz size differential shifts to the four-photon resonance at these beam powers~\cite{shiftsInPrep}. In the simulations mentioned above we adjust the $R_\perp$ beam polarization according to Figure~\ref{fig:differential_shift} to null the differential shift as effectively as possible.  

\begin{figure}[h]
\centering
    \hspace*{-0.0in}
    \includegraphics[width = 1.0\columnwidth]{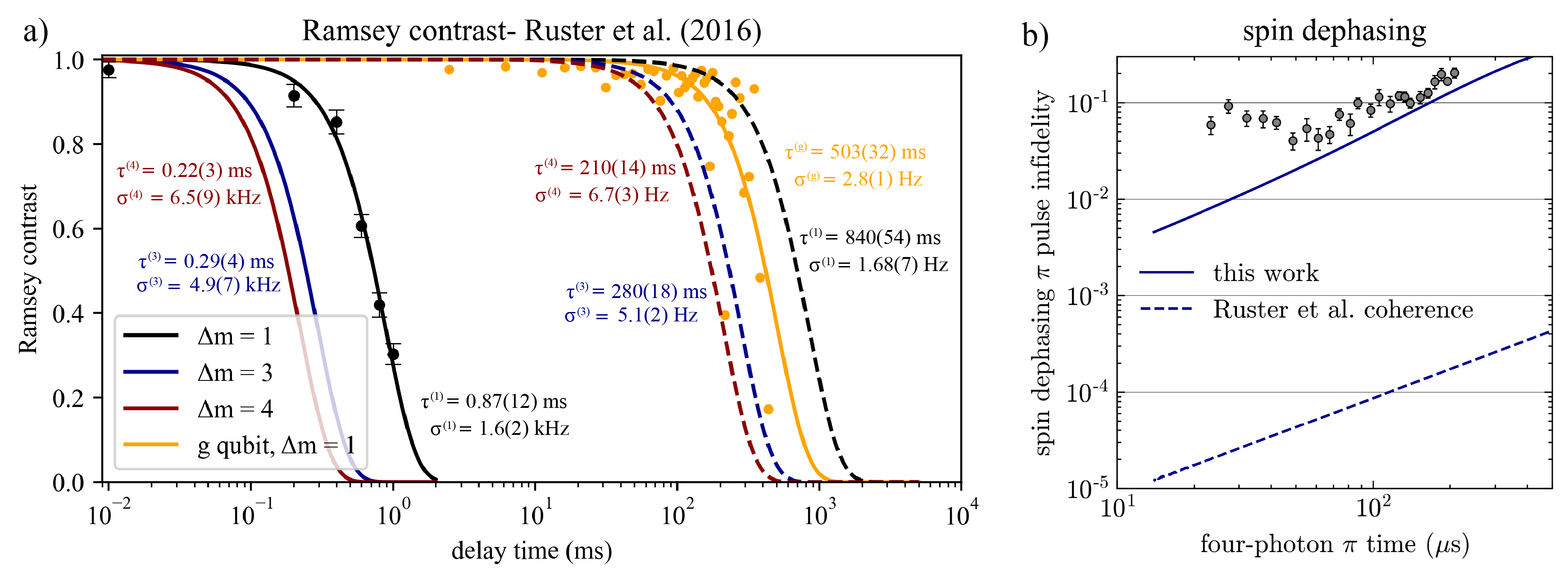}
    \caption{\textbf{a)}(black circles) Ramsey contrast measured in the $\mathrm{D_{5/2}}$ manifold in our experimental setup, with colored curves representing extrapolated coherences for larger $\Delta\mathrm{m}$ transitions within $\mathrm{D_{5/2}}$. (orange circles) Ramsey contrast presented in Ref.~\cite{coherenceImprovement} on a spin qubit in $\mathrm{S_{1/2}}$, with dashed curves corresponding to expected contrast for transitions in the $\mathrm{D_{5/2}}$, assuming the same noise sensitivity. \textbf{b} Four-photon $\pi$ pulse transfer infidelity due to spin dephasing, with magnetic field sensitivity in our experiment (solid) and the expected contribution assuming the experimental conditions of~\cite{coherenceImprovement} (dashed).}
\label{rustercoherence}
\end{figure}

\textbf{\textit{Spin decoherence}}
\newline
\label{decoherencesub}
We point to a demonstration of long Zeeman qubit spin coherence in the $\mathrm{S_{1/2}}$ manifold of a single trapped and laser cooled $^{40}$Ca$^+$ ion~\cite{coherenceImprovement} as an example of achievable performance in state-of-the-art systems with $\mathrm{Sm_2Co_{17}}$ permanent magnets supplying a 0.37\,mT quantizing magnetic field at the ion and a two-layer \textmu-metal shield to protect against ambient magnetic field fluctuations. Shown in Fig.~\ref{rustercoherence} is the Ramsey contrast as a function of delay time carried out in our experimental setup (black circles) and in the state-of-the-art setup~\cite{coherenceImprovement} (orange). The ratio of sensitivities to magnetic field fluctuations between the ground state qubit and a $\Delta\mathrm{m} = 1$ qubit encoded in the $\mathrm{D_{5/2}}$ manifold is given by the ratio of land\'e g factors of the two manifolds. Accordingly, a $\Delta\mathrm{m} = 1$ qubit in the $\mathrm{D_{5/2}}$ manifold has $3/5\times$ the magnetic field sensitivity to that of the ground qubit, corresponding to an expected $1/e$ coherence time of 850\,ms. We account for the increased magnetic field sensitivity of the higher $\Delta\mathrm{m}$ transitions within $\mathrm{D_{5/2}}$ as we do in Appendix~\ref{appendix:decoherence} and model decoherence in the same way, with the widths of shot-to-shot qubit errors extrapolated from the predicted coherence time. We expect an improvement of the $\Delta\mathrm{m}= 3$ four-photon transition coherence time of $\times 966$ in the apparatus used in Ref.~\cite{coherenceImprovement}. Similarly, we adjust the damping constant by that same factor to find $1/\gamma = 578$\,ms.

\begin{figure}[h]
\centering
    \hspace*{-0.0in}
    \includegraphics[width = 0.5\columnwidth]{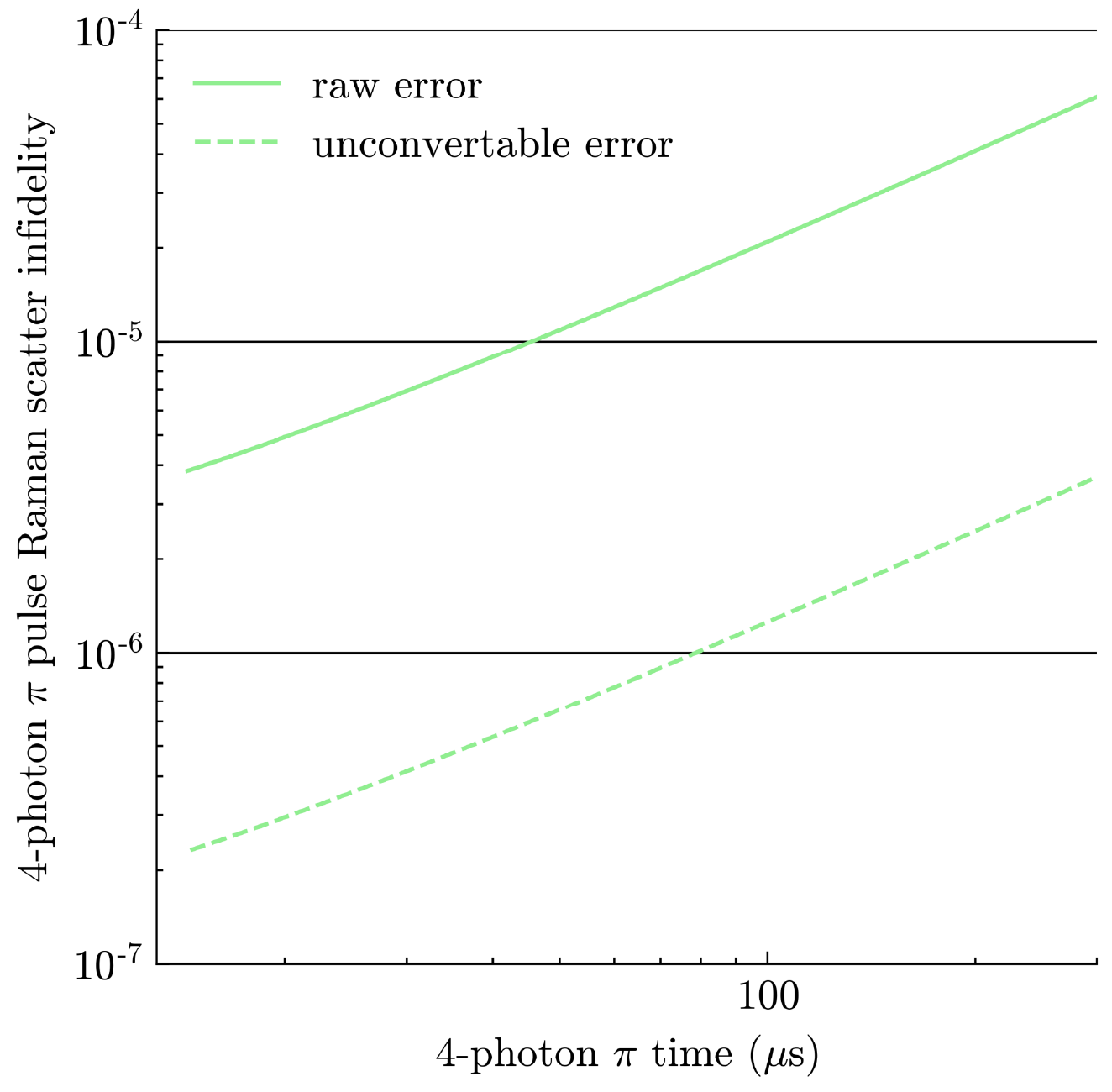}
    \caption{Infidelity in four-photon $\pi$ pulse from Raman scatter errorss (solid). Dashed lines correspond to Raman scatter errors that cannot be converted to an erasure error.} 
\label{fig:ramanscattering}
\end{figure}

\textbf{\textit{Spontaneous Raman scattering}}
\newline
\label{ramanscatteringsub}
We estimate contributions to the $\pi$ pulse transfer infidelity incurred by spontaneous Raman scattering via the $\mathrm{P_{3/2}}$ manifold according to the model presented in~\cite{danielScattering} and verified in~\cite{UCLAscattering, UOscattering}. For a $\pi$ pulse we assume population spends equal time in each qubit state, and for the four-photon transition presented in the main text the incurred infidelity is
\begin{align}
    \epsilon_{srs}=\frac{\Gamma_0+ \Gamma_3} {2}t_g
\end{align}
where $\Gamma_i$ is the scattering rate out of the state $\ket{i}$ as labeled in Fig.~\ref{fig:paths_flopping}b and $t_g$ is the $\pi$ time of the transition. We use the beam waist and polarization parameters given in Table~\ref{tab:fit_params}. In Fig.~\ref{fig:ramanscattering}a we plot the probability that a Raman scattering error occurred during the $\pi$ pulse at different $\pi$ times, corresponding to different $R_\perp$ beam power. From branching ratios $\sim$\,94\% of scattering events leave the qu$d$it manifold and scatter to either the $\mathrm{S_{1/2}}$ or $\mathrm{D_{3/2}}$ manifolds, both of which are bright under our fluorescence checks and can be detected and converted to an erasure error. The dashed line in Fig~\ref{fig:ramanscattering} corresponds to the probability of a Raman scatter error occurring that cannot be converted into an erasure error, where population has scattered back into the qudit subspace in the $\mathrm{D_{5/2}}$ manifold. 

\end{appendices}

\bibliography{refs}


\end{document}